\documentclass[aps,prb,twocolumn,showpacs,apsrev]{revtex4-1}
\usepackage[T1]{fontenc}
\usepackage{lmodern}
\usepackage[latin1]{inputenc}
\usepackage[swedish,english]{babel}
\usepackage{amsmath}
\usepackage{amsfonts}
\usepackage{graphicx} 
\usepackage{comment}
\usepackage{caption}
\usepackage{subcaption}
\usepackage{xfrac}
\usepackage{color}
\usepackage{multirow}
\usepackage{hyperref}
\usepackage{braket}
\usepackage{braket}

\hypersetup{
	colorlinks = true,
	pdftitle = {},
	pdfauthor = {},
	pdfkeywords = {},
	linkcolor = blue,
	citecolor = blue,
	filecolor = black,
	urlcolor = magenta
}

\newcommand{\img}{\mathrm{i}}

\graphicspath{{./graphics/}}

\begin{document}

\title{Magnetocrystalline anisotropy of Laves phase Fe$_2$Ta$_{1-x}$W$_x$ from first principles - the effect of 3d-5d hybridisation}
 
\author{Alexander Edstr\"om}
\affiliation{Department of Physics and Astronomy, Uppsala University, Box 516, 75121 Uppsala, Sweden}


\begin{abstract}
The magnetic properties of Fe$_2$Ta and Fe$_2$W in the hexagonal Laves phase are computed using density functional theory in the generalised gradient approximation, with the full potential linearised augmented plane wave method. The alloy Fe$_2$Ta$_{1-x}$W$_x$ is studied using the virtual crystal approximation to treat disorder. Fe$_2$Ta is found to be ferromagnetic with a saturation magnetization of $\mu_0 M_\text{s} = 0.66~\mathrm{T}$ while, in contrast to earlier computational work, Fe$_2$W is found to be ferrimagnetic with $\mu_0 M_\text{s} = 0.35~\mathrm{T}$. The transition from the ferri- to the ferromagnetic state occurs for $x \leq 0.1$. The magnetocrystalline anisotropy energy (MAE) is calculated to $1.25~\mathrm{MJ/m^3}$ for Fe$_2$Ta and $0.87~\mathrm{MJ/m^3}$ for Fe$_2$W. The MAE is found to be smaller for all values $x$ in Fe$_2$Ta$_{1-x}$W$_x$ than for the end compounds and it is negative (in-plane anisotropy) for $0.1 \leq x \leq 0.9$. The MAE is carefully analysed in terms of the electronic structure. Even though there are weak 5d contributions to the density of states at the Fermi energy in both end compounds, a reciprocal space analysis, using the magnetic force theorem, reveals that the MAE originates mainly from regions of the Brillouin zone with strong 3d-5d hybridisation near the Fermi energy. Perturbation theory and its applicability in relation to the MAE is discussed. 
\end{abstract}

\maketitle

The magnetocrystalline anisotropy energy (MAE) is the intrinsic relativistic feature, originating from spin-orbit coupling (SOC)\cite{Vleck1937}, of magnetic materials that the energy depends on the direction of magnetization relative to the crystal lattice. It is crucial in a wide range of applications, from permanent magnets\cite{Gutfleisch2011, Niarchos2015, Coey2011, Hirosawa2015} to magnetic storage devices\cite{Weller2000}. The SOC is strong in heavy elements such as rare-earths (REs) and actinides which consequently acquire large MAE, while in applications it is highly desirable to obtain a large MAE without such expensive or inaccessible constituent elements\cite{Kramer2010}. One compound which has gained much attention due to its huge MAE is tetragonal FePt\cite{Staunton2004,Staunton2004a,Okamoto2002,Burkert2005a,Lyubina2005}. This material acquires its magnetisation mainly from Fe, while the important factors resulting in the large MAE include the strong SOC of the Pt atom, as well as the uniaxial crystal structure. The crystal structure is crucial because highly symmetric, e.g. cubic, crystals tend to have at least one order of magnitude lower MAE. Nevertheless, FePt contains large amounts of the valuable element Pt, whereby alternative magnetic 3d-5d composites in uniaxial crystal structures can be of great technological value. One such compound is hexagonal Laves phase Fe$_2$W, which was initially reported by Arnfelt and Westgren\cite{Arnfelt1935} and recently attracted some attention in the context of permanent magnet replacement materials\cite{Kumar2014, Koten2015}. Early electronic structure calculations\cite{Ishida1985} failed to establish the existence of ferromagnetism in the compound from the Stoner criterion. While it now seems clear that the compound is magnetically ordered\cite{Kumar2014, Koten2015}, a thorough understanding of the magnetism in this material appears to be absent in literature and some discrepancies can be seen between recent computational\cite{Kumar2014} and experimental work\cite{Koten2015}. For example, calculations\cite{Kumar2014} overestimated the saturation magnetization by nearly thirty percent and provided a vastly different MAE when compared to experimental data from nanoparticles\cite{Koten2015}. It is therefore the purpose of this work to use state of the art electronic structure calculations to unambiguously determine the magnetic ground state of the Fe$_2$W compound and investigate the magnetic properties, including the technologically important intrinsic properties of saturation magnetization ($M_\text{s}$) and MAE. The closely related compound Fe$_2$Ta is isostructural to Fe$_2$W\cite{BLAZINA1989247} and also studied. Some focus will be put on the MAE, which will be carefully analysed in terms of the electronic structure. Furthermore, the possibility to tune the MAE by alloying W and Ta will be examined and a discussion of the underlying physical principles provided. 

Density functional theory (DFT) calculations in the generalized gradient approximation\cite{PhysRevLett.77.3865} (GGA) were performed with the full-potential linearized augmented plane waves (FP-LAPW) method as implemented in WIEN2k\cite{Blaha2001}. Initially, spin-polarized calculations were performed in the scalar relativistic approximation, but to calculate the MAE, SOC must be included and this was done in a second variational approach\cite{Koelling1977}. The size of the basis set used is typically described by the product of the smallest muffin-tin sphere and the largest reciprocal lattice vector included, $RK_\text{max}$. For structure optimizations, this value was set to $RK_\text{max}=7$, while for MAE calculations a larger value of $RK_\text{max}=9$ was used. To obtain a well converged formation energy, a value as large as $RK_\text{max}=9.5$ was needed. Integration of $\mathbf{k}$-points over the Brillouin zone was performed using the improved tetrahedron method\cite{Blochl1994} and 700 $\mathbf{k}$-points in the full Brillouin zone (48 in the irreducible wedge of the Brillouin zone after considering the 24 symmetry operations of the crystal) were used for structure optimization, 1500 for the calculation of formation energy and as many as 30000 $\mathbf{k}$-points were used in order to obtain well converged MAE values. 

One unit cell of the relevant crystal structure contains two inequivalent Fe positions with multiplicity two and six respectively, as well as two equivalent 5d sites. Calculations were performed with the initial spin state either in ferro or ferrimagnetic configurations, i.e. parallel or antiparallel alignment of spins on the two different Fe positions. In the case of Fe$_2$Ta, the total energy was found to be approximately 1.8 eV per unit cell lower in the case of ferromagnetic ordering compared to ferrimagnetic ordering.  For Fe$_2$W, on the other hand, all calculations converged into the ferrimagnetic state, regardless of initial spin configuration and lattice parameters. Lattice parameters were calculated by minimizing the total energy with respect to volume and $c/a$ and relaxing the internal atomic positions in each step. The calculated lattice parameters are reported in Table~\ref{table1}, which also contains spin magnetic moments and the corresponding saturation magnetizations as well as formation energies. For Fe$_2$Ta, the lattice parameters have been experimentally reported as $a=4.833~\text{\AA}$ and $c=7.868~\text{\AA}$~\cite{BLAZINA1989247} and for Fe$_2$W, $a=4.727~\text{\AA}$ and $c=7.704~\text{\AA}$~\cite{Arnfelt1935} , in close agreement with the calculated values in Table~\ref{table1}, although for Fe$_2$W, $c/a$ is slightly larger in the calculated data. The Fe moments in Fe$_2$W are of similar size and opposite sign but as there are two and six of the respective Fe sites in one unit cell, there is a net total of $4.45\mu_\text{B}\text{/u.c.}$, corresponding to a saturation magnetization of $\mu_0 M_\text{s} = 0.35~\text{T}$. Since in Fe$_2$Ta the Fe moments are parallel, the total magnetic moment and corresponding saturation magnetization is significantly larger, reaching a value of $\mu_0 M_\text{s} = 0.66~\text{T}$. Ta and W have a small induced moments of $-0.24\mu_\text{B}$ and $-0.05\mu_\text{B}$, anti-parallel to the total magnetic moment, respectively, as is typical for these 5d atoms in a magnetic 3d host\cite{Wienke1991}.
\begin{table}
\caption{\label{table1} Lattice parameters and parameters of the internal atomic positions, magnetic moments, saturation magnetization and formation energy of Fe2W and Fe2Ta as calculated in a scalar relativistic, spin polarized GGA calculation, neglecting SOC, in WIEN2k. }
\begin{ruledtabular}
\begin{tabular}{l r r}
  & Fe$_2$Ta & Fe$_2$W \\
\hline  
a (\AA) & 4.811 & 4.674  \\
c (\AA) &  7.874 & 7.768 \\
$x_\mathrm{Fe_2}$ & 0.83192 & 0.82946 \\
$z_\text{5d}$ & 0.06405 & 0.06924\\
$m(\text{Fe}_1)~(\mu_\text{B})$ & 0.90  & -1.14 \\
$m(\text{Fe}_2)~(\mu_\text{B})$ & 1.43 & 1.17 \\
$m(\text{5d})~(\mu_\text{B})$ & -0.24  & -0.05 \\ 
$m_\text{tot}~(\mu_\text{B}/\text{u.c.})$ & 8.88 & 4.45  \\ 
$\mu_0 M_\text{s}$ (T) & 0.66 & 0.35 \\
Formation energy (eV/u.c.) & -2.82 & -0.63 \\ 
\end{tabular}
\end{ruledtabular}
\end{table}

Since a different magnetic ordering, with a magnetic moment close to zero on the first Fe site and a larger moment moment around $1.3\mu_\text{B}$ on the second Fe atom, has been reported in earlier computational work (pseudopotential DFT calculations in the GGA)\cite{Kumar2014} for Fe$_2$W, further investigation seems necessary to unambiguously determine the correct magnetic ground state within the GGA. Hence, fixed spin moment calculations, allowing the total magnetic moment of the system to be constrained to a fixed given value, were performed. The total magnetic moment was varied around the value of $6.8\mu_\text{B}/\text{u.c.}$, previously reported\cite{Kumar2014}. Magnetic moments of $-0.05\mu_\text{B}$ and $1.25\mu_\text{B}$ were then obtained on the two Fe atoms, which is similar to the earlier computational results\cite{Kumar2014}. Initially, the lattice parameters were set to the values mentioned in Ref~\cite{Kumar2014} but then attempts were made at optimizing the crystal structure with fixed magnetic moment to lower the energy further. However, all calculations resulted in total energies which were higher than those obtained for the structure given in Table~\ref{table1} and no minimum could be located in the total energy as function of total magnetic moment. Based on these results, the most probable conclusion appears to be that the authors of Ref~\cite{Kumar2014} assumed a ferromagnetic order as initial state and reached a local energy minimum for the magnetic moments were reported. The correct magnetic moments corresponding to the global energy minimum, within the GGA, based on all results obtained here, are expected to be those in Table~\ref{table1}. The explanation given here is consistent with the observation that the previous computational work presented a value of $\mu_0 M_\text{s}$ as approximately $0.56~\text{T}$ which overestimated the experimental low temperature value of approximately $\mu_0 M_\text{s} = 0.44~\text{T}$. Nevertheless, the value given in this work somewhat underestimates the experimental result. A possible source of discrepancy is surface effects of the nanoparticles, where enhanced magnetic moments could appear near the surface. 

Somewhat surprisingly, a non-negligible difference is seen also in lattice parameters and total magnetic moment for Fe$_2$Ta, when comparing to previous computational work~\cite{Kumar2014}, where $a=4.825~\AA$ and $c/a=1.6329$ (corresponding to $c=7.879~\AA$) was reported. The difference in $a$ is merely 2\% and might be expected for the two different computational methods. The difference in total spin magnetic moment is, however, larger. For example, the magnetic moment on the Fe$_1$ is computed to $0.90~\mu_\text{B}$, while the other authors reported a value well above $1~\mu_\text{B}$. The reason for this discrepancy is difficult to pinpoint exactly, as both sets of calculations are performed in the GGA\cite{PhysRevLett.77.3865}, but might partly be related to the difference in lattice parameters.

By comparing the total energy of Fe$_2$(Ta/W) in the calculated ground state with that of bcc Fe and bcc Ta or W, the formation energy was calculated to $-0.63~\text{eV/u.c}$ for Fe$_2$W, which is lower than the value  close to zero previously reported\cite{Kumar2014}. A negative formation energy is expected for a stable phase and a possible scenario appears to be that the authors of Ref.~\cite{Kumar2014} obtained a too high formation energy due to calculating a local energy minimum and thus a too high total energy. For Fe$_2$Ta, the formation energy is lower and this compound may therefore be expected to be more stable and form more easily in nature. 

In order to compute the MAE, calculations were performed including SOC, which also results in a non-zero orbital magnetic moment, that is otherwise quenched. The computed spin magnetic moments ($m_\text{S}$), orbital magnetic moments ($m_\text{L}$) and MAEs are listed in Table~\ref{table2}. When the magnetization is along the $a$-axis, the SOC results in a lowering of symmetry so that the second Fe site with initially six equivalent atoms are split into two types with two and four Fe atoms of each, labelled Fe$_2$ and Fe$_3$ respectively. Hence, the spin and orbital moments are same for Fe$_2$ and Fe$_3$ when the magnetization is along the $c$-axis but not when it is along the $a$-axis. The MAE is calculated to $E_\text{MAE} = 1.24~\text{meV/u.c.} = 1.25~\text{MJ/m}^3$ for Fe$_2$Ta and $E_\text{MAE} = 0.79~\text{meV/u.c.} = 0.87~\text{MJ/m}^3$ for Fe$_2$W, with easy magnetization axis along the $c$-direction of the crystal in both cases. The calculated uniaxial MAE for Fe$_2$W presented here is in better agreement with the small uniaxial MAE recently presented in experimental work\cite{Koten2015} than the large in plane MAE previously computed for the Fe$_2$W compound\cite{Kumar2014}. Nevertheless, the computed value found in this work is significantly larger than the reported experimental value of $286~\text{kerg/cm}^3 = 28.6~\text{kJ/m}^3$. However, measurements have only been presented for nanoparticles, while unambiguous MAE measurements require single crystals. For Fe$_2$Ta, an experimental MAE has not been found in literature, but the value calculated here differs notably from the value of $E_\text{MAE} = -1.4~\text{meV/u.c.}$ previously calculated~\cite{Kumar2014}. This discrepancy is most likely related to the difference in magnetic moments obtained, as mentioned above, but could also be partially related to other computational details, such as the treatment of SOC or core electrons. 
\begin{table}
\caption{\label{table2} Spin magnetic moments, $m_\text{S}$, orbital magnetic moments $m_\text{L}$, saturation magnetizations and MAE for Fe$_2$W as calculated in WIEN2k, including SOC with magnetization either along 100 or 001 directions and using the lattice parameters presented in Table \ref{table1}. }
\begin{ruledtabular}
\begin{tabular}{l r r}
Fe$_2$Ta & $\mathbf{m} \parallel 100 $ & $\mathbf{m} \parallel 001 $\\
\hline
$m_S(\text{Fe}_1)~(\mu_\text{B})$ 	&0.943	& 0.932	 \\
$m_S(\text{Fe}_2)~(\mu_\text{B})$ 	&1.433	& 1.432 	 \\
$m_S(\text{Fe}_3)~(\mu_\text{B})$ 	&1.427  	& 1.432	 \\
$m_S(\text{Ta})~(\mu_\text{B})$ 		& -0.240	& -0.238 	 \\ 
$m_L(\text{Fe}_1)~(\mu_\text{B})$ 	& 0.070	& 0.109 	 \\
$m_L(\text{Fe}_2)~(\mu_\text{B})$ 	& 0.091	& 0.099 	 \\
$m_L(\text{Fe}_3)~(\mu_\text{B})$ 	& 0.101	& 0.099	 \\
$m_L(\text{Ta})~(\mu_\text{B})$ 		& 0.033 	& 0.034 	 \\ 
$\mu_0 M_\text{s}$ (T) 										& 0.69		& 0.69	 \\
Energy (meV/u.c.) 									& 1.24 	& 0 		 \\
Energy (MJ/m$^3$) 								& 1.25 	& 0 		 \\
\hline 
\hline 
Fe$_2$W & $\mathbf{m} \parallel 100 $ & $\mathbf{m} \parallel 001 $\\
\hline
$m_S(\text{Fe}_1)~(\mu_\text{B})$ 	& -1.148 		& -1.150 	\\
$m_S(\text{Fe}_2)~(\mu_\text{B})$ 	& 1.163  		& 1.172 	\\
$m_S(\text{Fe}_3)~(\mu_\text{B})$ 	& 1.172  		& 1.172 	\\
$m_S(\text{W})~(\mu_\text{B})$ 		& -0.044		& -0.044 	\\ 
$m_L(\text{Fe}_1)~(\mu_\text{B})$ 	& -0.066 		& -0.151 	\\
$m_L(\text{Fe}_2)~(\mu_\text{B})$ 	& 0.045 		& 0.039 	\\
$m_L(\text{Fe}_3)~(\mu_\text{B})$ 	& 0.074 		& 0.039 	\\
$m_L(\text{W})~(\mu_\text{B})$ 		& 0.002 		& 0.002 	\\ 
$\mu_0 M_\text{s}$ (T) 										& 0.36 		& 0.35 	\\
Energy (meV/u.c.) 									& 0.79 		& 0 		\\
Energy (MJ/m$^3$) 								& 0.87 		& 0 		\\
\end{tabular}
\end{ruledtabular}
\end{table}

\begin{figure}[hbt!]
	\centering
	\includegraphics[width=0.49\textwidth]{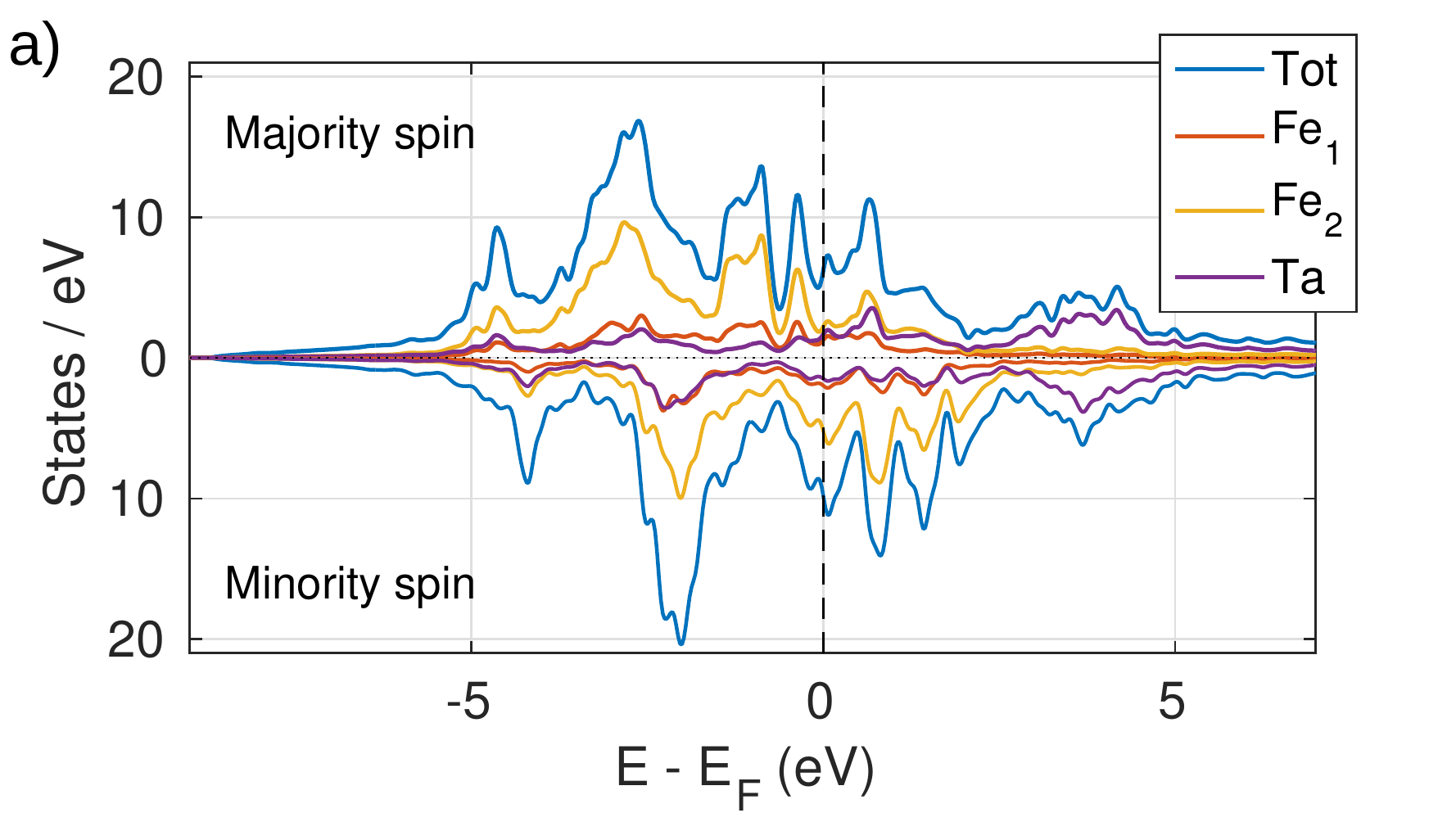} \\
	\includegraphics[width=0.49\textwidth]{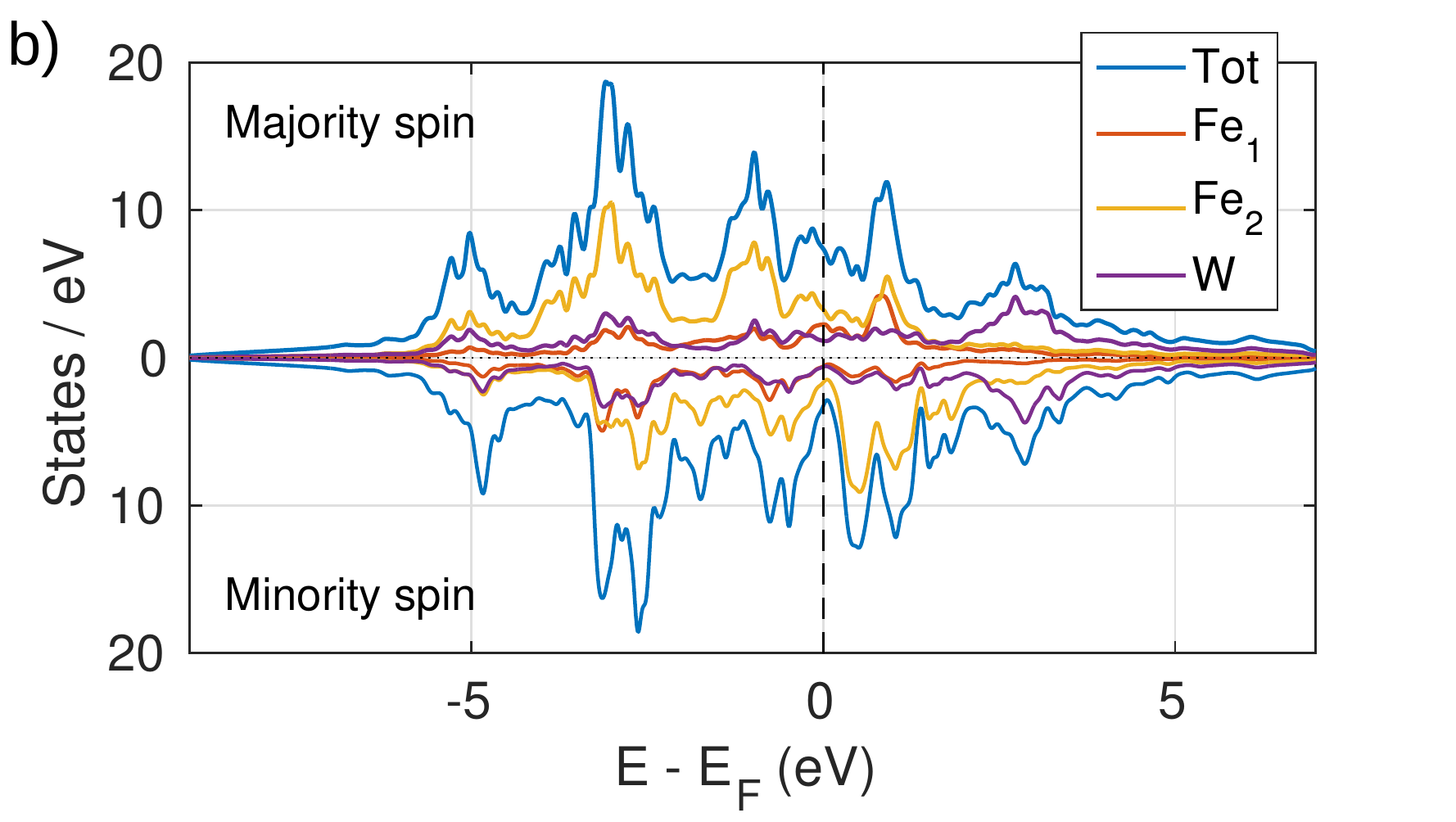}
	\caption{Spin polarized DOS  for Fe$_2$Ta in a) and Fe$_2$W in b).}
	\label{dos}
\end{figure}

Fig.~\ref{dos} shows the spin polarized density of states (DOS) for Fe$_2$Ta (a) and Fe$_2$W (b). The majority spin DOS is similar for the two compounds, with the Fermi energy ($E_\text{F}$) at approximately the same location. However, as Ta is exchanged for W more electrons are added into the system and the minority spin states become occupied, whereby these are shifted more to the left in Fig.~\ref{dos}b) and, as a result, $E_\text{F}$ coincides with the bottom of a valley in the minority spin DOS of Fe$_2$W. Thus the DOS($E_\text{F}$) for Fe$_2$W is dominated by minority spin states, in contrast to Fe$_2$Ta, where the opposite is true. This fact will be of importance later when analysing the relation between MAE and orbital moment anisotropy. It is also interesting to note that the minority spin DOS of Fe$_2$W has a valley at $E_\text{F}$, resulting in a higher degree of spin polarization of the DOS($E_\text{F}$) compared to Fe$_2$Ta. In both cases, the DOS($E_\text{F}$) is dominated by Fe, with rather modest contributions from the 5d atoms. This might be one important reason, together with other details in the band structure around $E_\text{F}$, why these compounds do not possess larger MAE. Even the L1$_0$ phase of MnAl exhibits an MAE well above $1~\text{MJ/m}^3$\cite{Edstrom2014} without any constituent element heavier than a 3d atom. Heavier atoms, such as 5d's, should allow significantly larger MAE, e.g., $4~\text{MJ/m}^3$~\cite{Okamoto2002} or more~\cite{Thiele1998} in FePt. However, this requires significant 3d-5d hybridisation around $E_\text{F}$, as is seen in FePt\cite{PhysRevB.63.144409}, but appears to be limited in the compounds studied here. Nevertheless, the contribution from Ta (3.2 states/eV for both spin channels summed) is greater than that of W (1.8 states/eV). This is consistent with the observation that the MAE is greater in the compound containing Ta, although other differences in the electronic structure are also expected to play a role. One more interesting observation in the DOS is that the minority spin DOS of Fe$_2$W has a valley at $E_\text{F}$, resulting in a higher degree of spin polarization of the DOS($E_\text{F}$) compared to Fe$_2$Ta.

In a system with weak SOC, such as 3d-based itinerant magnets, where $\xi$ is significantly smaller than the bandwidth (less than $100~\text{meV}$ compared to several eV), it is reasonable to describe the effect of SOC in terms of perturbation theory and important insights can be gained by doing so\cite{Kondorskii1973,Bruno1989,Andersson2007}. For a uniaxial crystal the leading term is of second order while for cubic crystals it is fourth order. Andersson \emph{et al.}\cite{Andersson2007} discussed the case of having several atomic types and hybridisation between these in a tight-binding description. One can consider unperturbed single particle states at the point $\mathbf{k}$ in the Brillouin zone as
\begin{equation}
\ket{\mathbf{k}, i} = \sum_{q,\mu} c_{\mathbf{k},i,q,\mu} \ket{\mathbf{k}, q, \mu, \sigma_i},
\label{multisite_orbs}
\end{equation}  
with summation over atomic sites $q$ and orbital states $\mu$, but not over the spin $\sigma_n$ since the unperturbed states each have well defined spin. With on site SOC, The shift in the energy eigenvalue $E_{\mathbf{k}, i}$ associated with $\ket{\mathbf{k}, i}$ is
\begin{align}
	&\Delta E_{\mathbf{k}, i}(\hat{\mathbf{n}}) = -\sum_{j \neq i } \sum_{q q^{\prime}} \sum_{\mu\mu^{\prime}\mu^{\prime\prime}\mu^{\prime\prime\prime}}
	n_{\mathbf{k},i,q\mu,q^{\prime}\mu^{\prime\prime\prime}} n_{\mathbf{k},j,q^{\prime}\mu^{\prime\prime},q\mu^{\prime}}  \nonumber \\ 
	 & \cdot \frac{ \langle q\mu\sigma_i | \xi_q \mathbf{\hat{l}} \cdot \mathbf{\hat{s}} | q\mu^{\prime} \sigma_j \rangle \langle  q^{\prime}\mu^{\prime \prime}\sigma_j | \xi_{q^{\prime}} \mathbf{\hat{l}} \cdot \mathbf{\hat{s}} |  q^{\prime}\mu^{\prime\prime\prime} \sigma_i \rangle }
	 {E_{\mathbf{k},j}-E_{\mathbf{k},i} } ,
	 \label{HybridisationSOC}
\end{align} 
with occupation numbers $n_{\mathbf{k},i,q\mu,q^{\prime}\mu^{\prime\prime\prime}} = c^\ast_{\mathbf{k},i,q,\mu}c_{\mathbf{k},i,q^\prime,\mu^{\prime\prime\prime}}$ and spin and orbital angular momentum operators $\mathbf{\hat{s}}$ and $\mathbf{\hat{l}}$. For a given $q$ and $\mathbf{k}$, it is clear that the effect of the SOC is determined by matrix elements of the form $\bra{\mu_i, \sigma_i}\mathbf{\hat{l}} \cdot \mathbf{\hat{s}}\ket{\mu_j, \sigma_j}$ and for convenience these are listed with respect to spin and d-orbitals in the appendix. $\hat{\mathbf{n}}$ is the spin quantization axis (magnetization direction) and the dependence of $\Delta E_{\mathbf{k}, i}(\hat{\mathbf{n}})$ on this quantity comes from the SOC matrix elements. For the total shift in $E_{\mathbf{k}, i}$, the coupling between all states $j \neq i$ should be considered. However, if both $i$ and $j$ denote occupied states there will be a cancellation when these are summed over to compute the total energy. Therefore, only coupling between occupied and unoccupied states are relevant, except possibly in the small regions of the Brilloiun zone where deformations of the Fermi surface occur, as was pointed out by Kondorskii and Straube\cite{Kondorskii1973}. This leads to the important and well established conclusion that the MAE is determined by the electronic band structure near the Fermi energy, in particular by the coupling between occupied and unoccupied states. One more important observation from Eq.~\ref{HybridisationSOC} is that regions in the band structure with significant Fe-Ta hybridisation will allow MAE contributions of order $\frac{\xi_\text{Ta}\xi_\text{Fe}}{E_{\mathbf{k},j}-E_{\mathbf{k},i}}$, which is significantly larger than $\frac{\xi_\text{Fe}^2}{E_{\mathbf{k},j}-E_{\mathbf{k},i}}$, since $\xi_\text{Ta}$ is several times larger than $\xi_\text{Fe}$, or similarly for W instead of Ta.

\begin{figure*}[hbt!]
	\centering
    \begin{subfigure}[b]{0.49\textwidth}
        \includegraphics[trim = 0mm 6.5mm 0mm 3mm, clip, width=\textwidth]{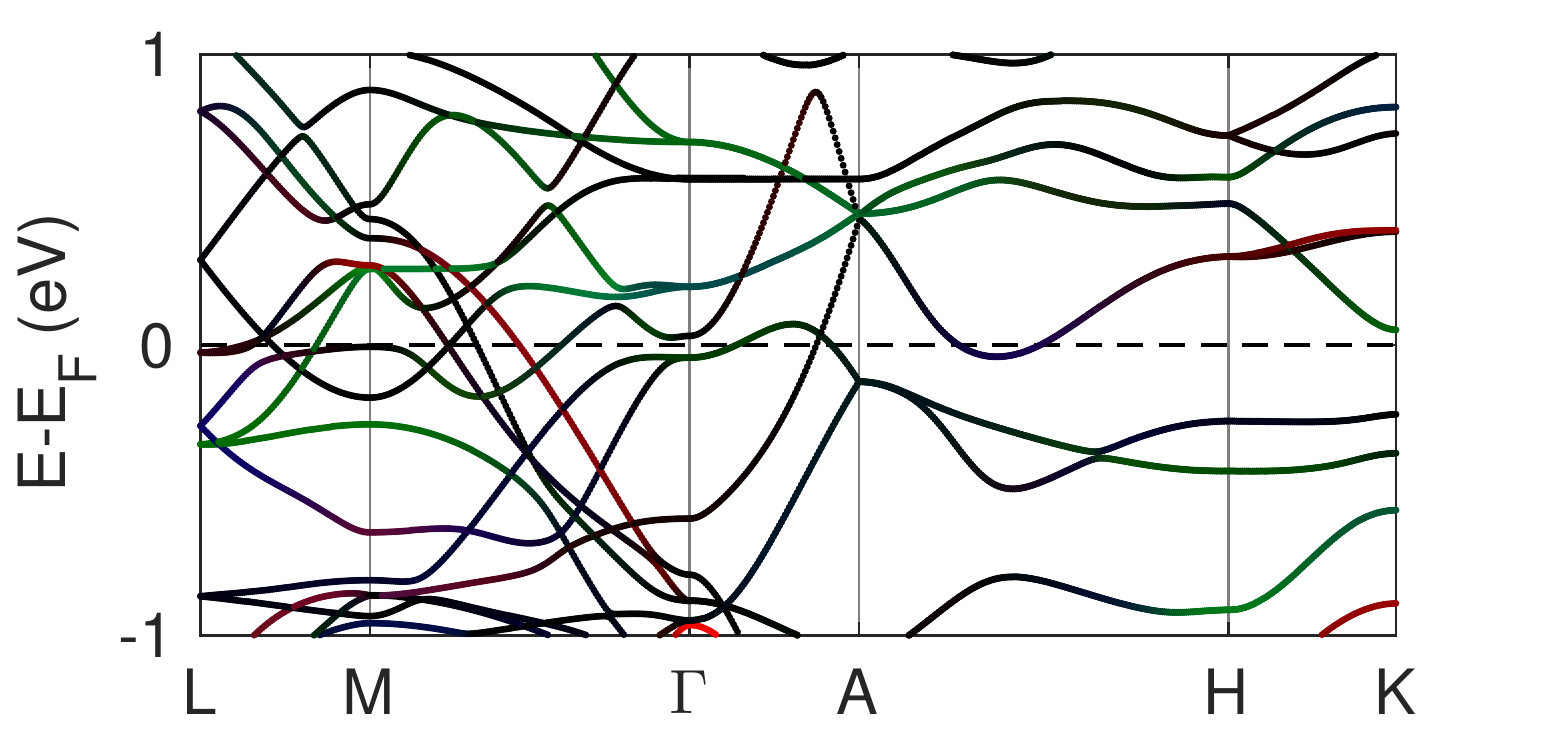}
        \caption{Fe$_1$, spin up. }
        \label{fig:Fe1_up}
    \end{subfigure}	
    \begin{subfigure}[b]{0.49\textwidth}
        \includegraphics[trim = 0mm 6.5mm 0mm 3mm, clip, width=\textwidth]{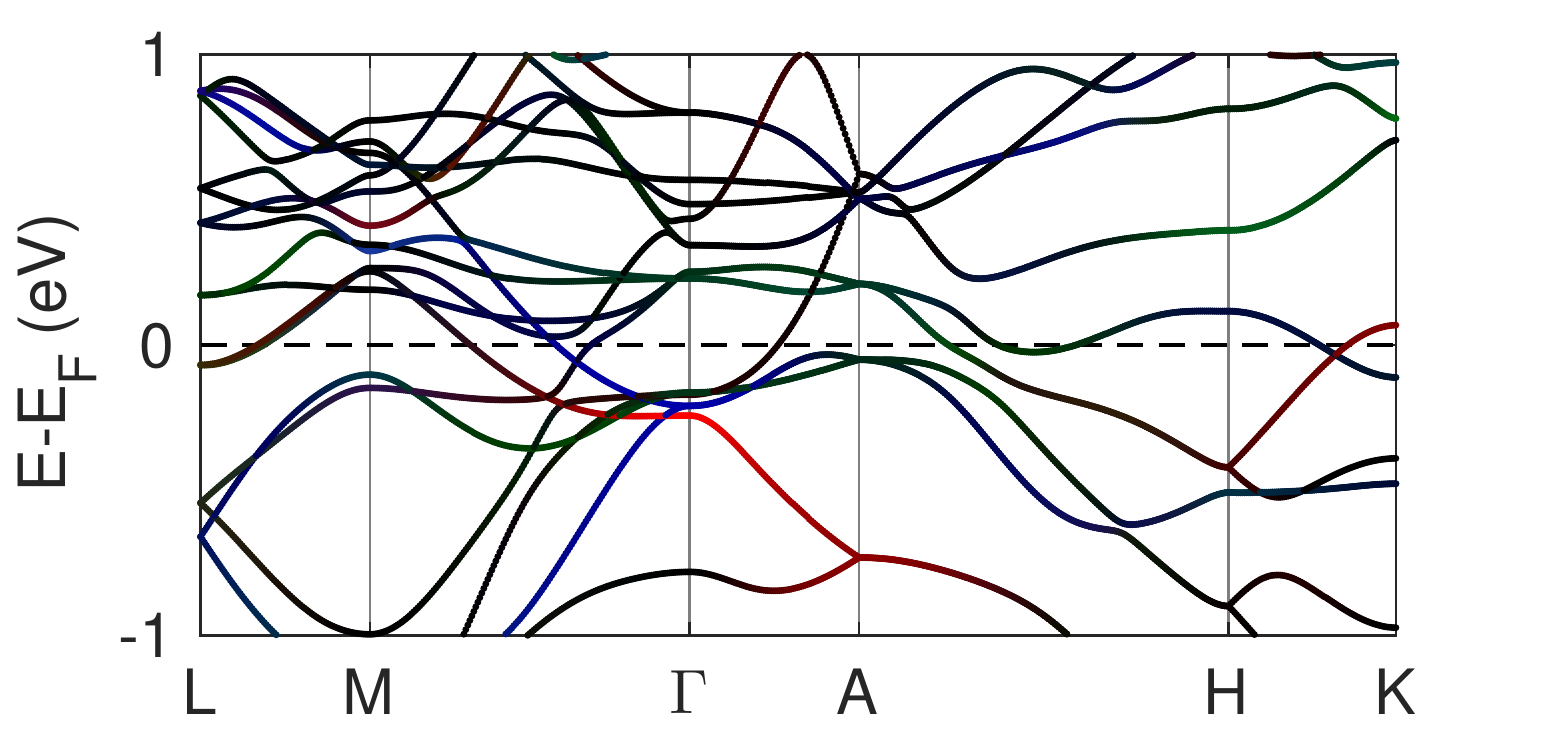}
        \caption{Fe$_1$, spin down. }
        \label{fig:Fe1_dn}
    \end{subfigure}	
    \begin{subfigure}[b]{0.49\textwidth}
        \includegraphics[trim = 0mm 6.5mm 0mm 3mm, clip, width=\textwidth]{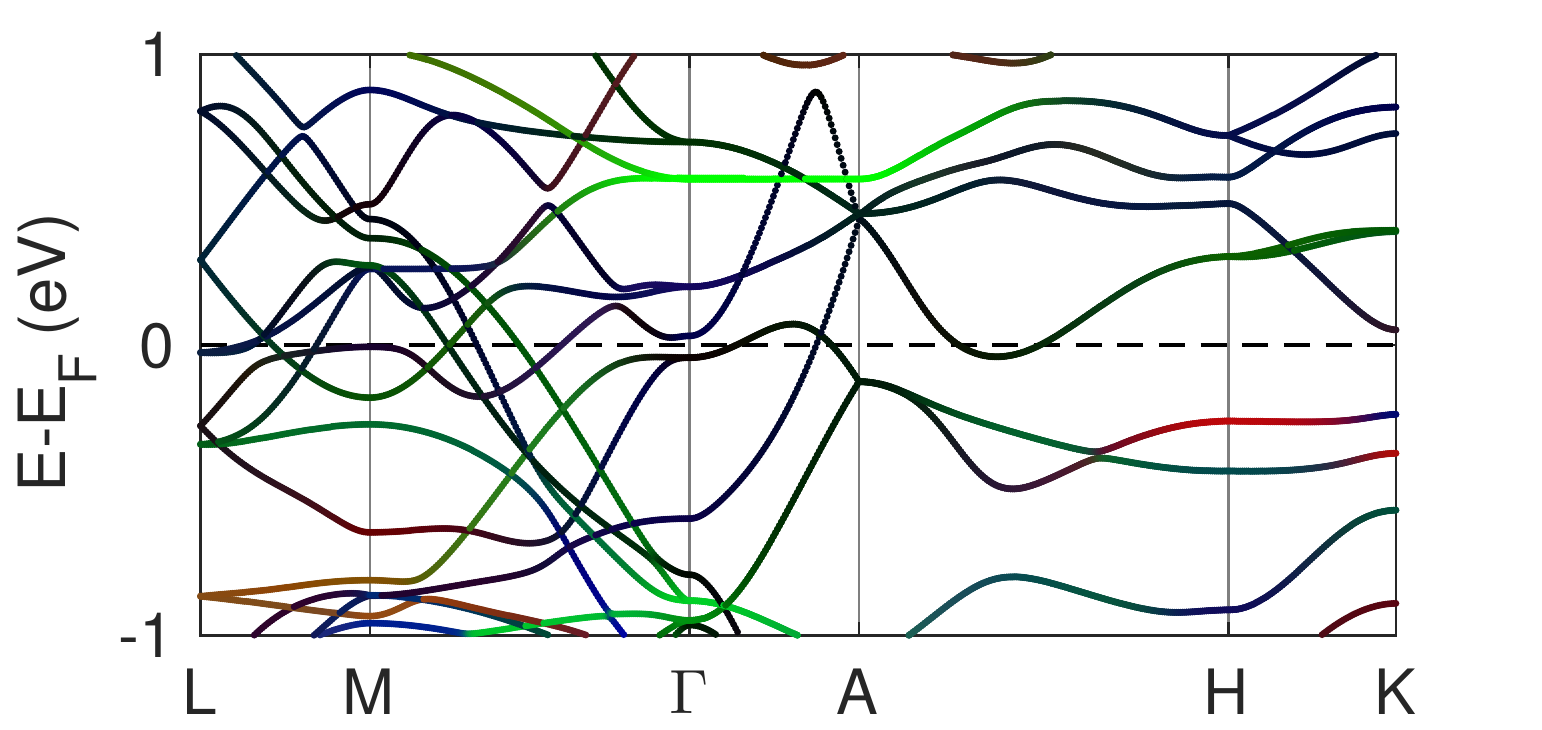}
        \caption{Fe$_2$, spin up. }
        \label{fig:Fe2_up}
    \end{subfigure}	
    \begin{subfigure}[b]{0.49\textwidth}
        \includegraphics[trim = 0mm 6.5mm 0mm 3mm, clip, width=\textwidth]{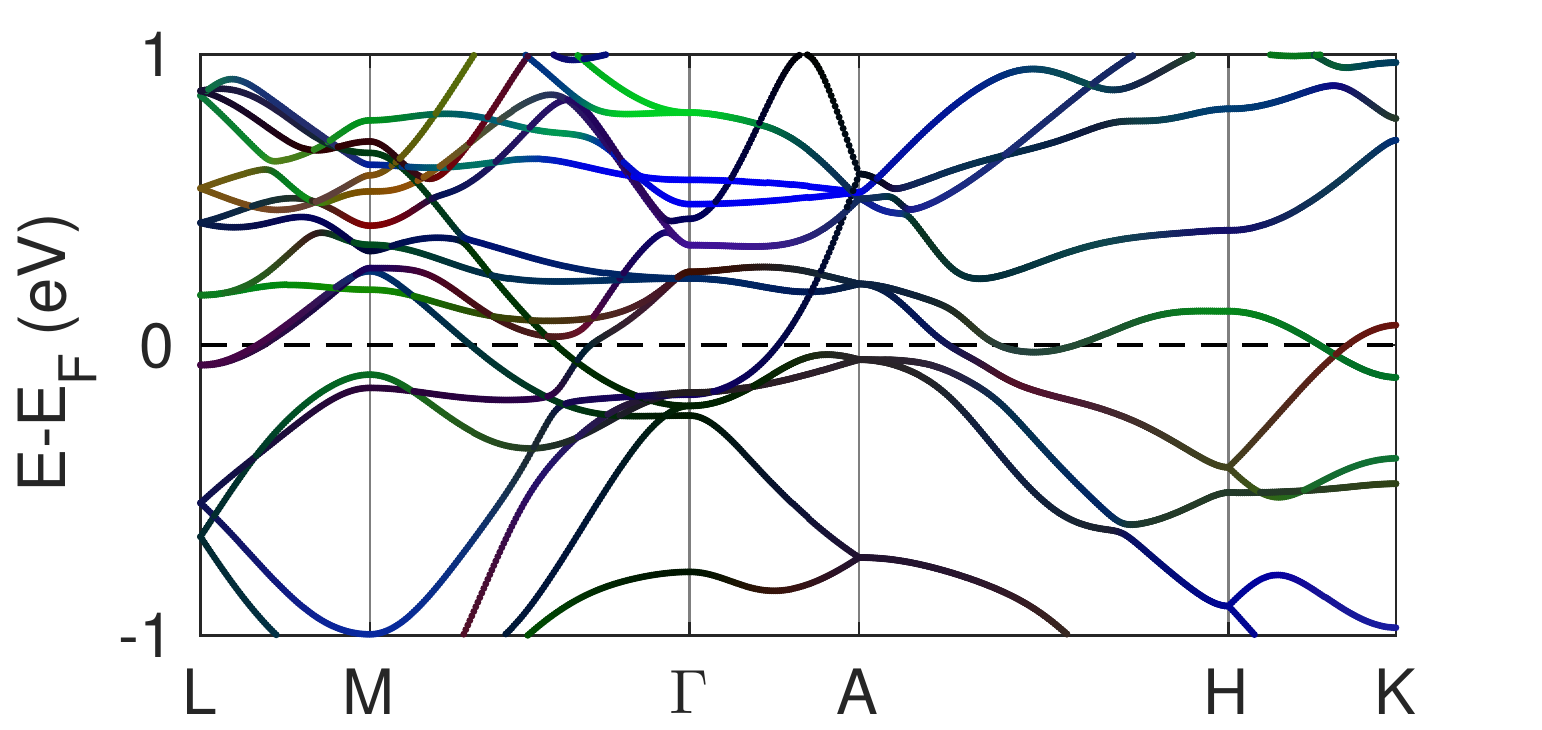}
        \caption{Fe$_2$, spin down. }
        \label{fig:Fe2_dn}
    \end{subfigure}	
    \begin{subfigure}[b]{0.49\textwidth}
        \includegraphics[trim = 0mm 1mm 0mm 3mm, clip, width=\textwidth]{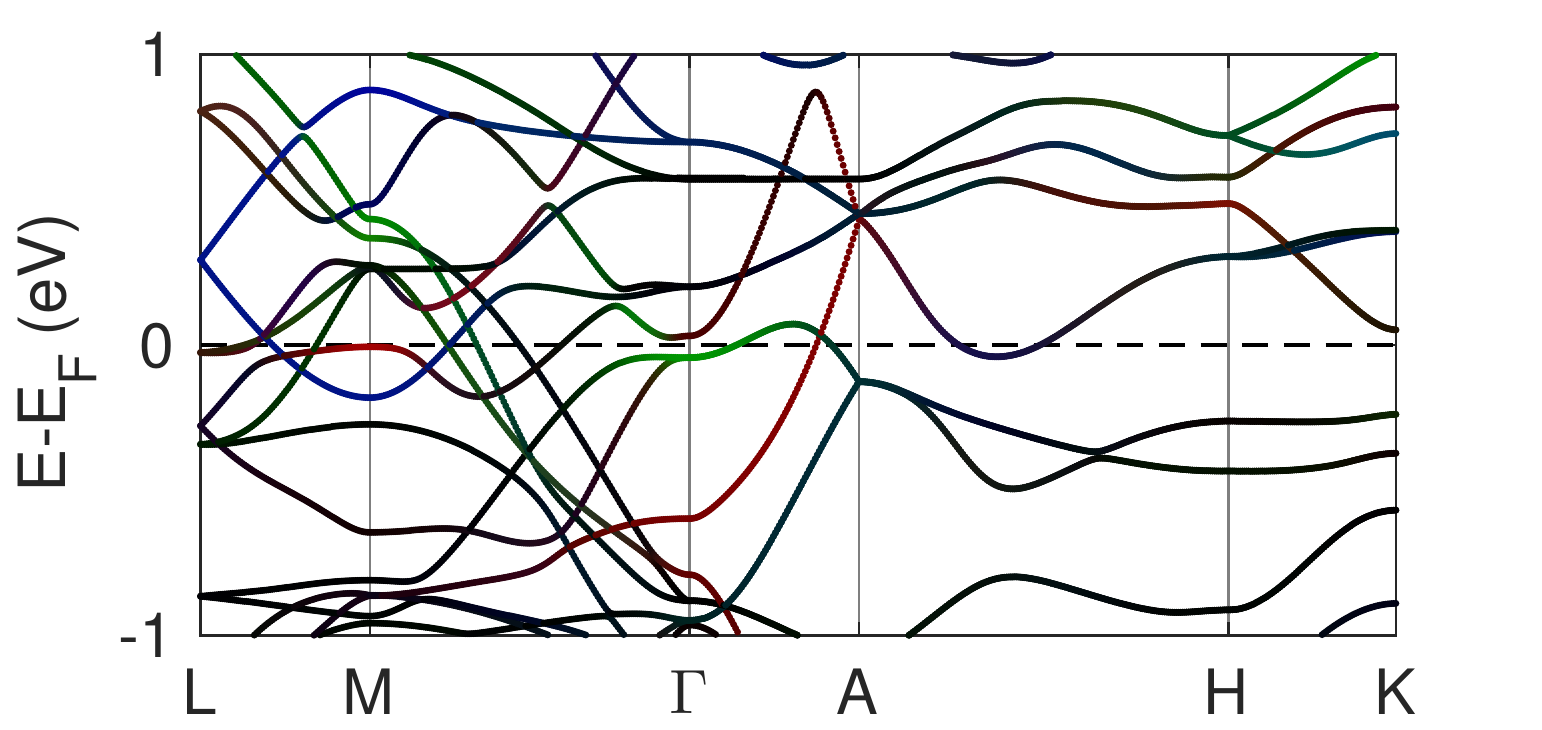}
        \caption{Ta, spin up. }
        \label{fig:Ta_up}
    \end{subfigure}	
    \begin{subfigure}[b]{0.49\textwidth}
        \includegraphics[trim = 0mm 1mm 0mm 3mm, clip, width=\textwidth]{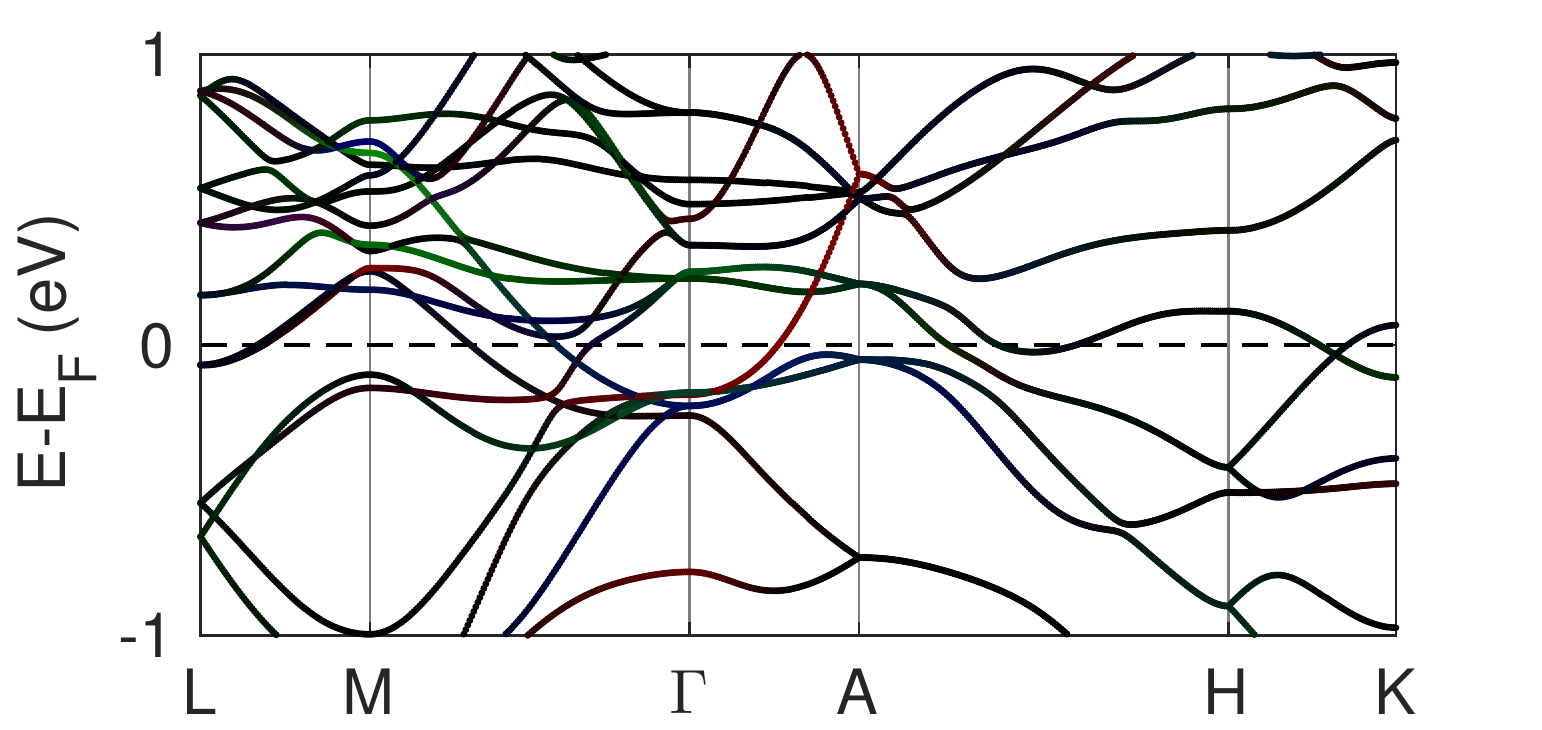}
        \caption{Ta, spin down. }
        \label{fig:Ta_dn}
    \end{subfigure}	
    \begin{subfigure}[b]{0.92\textwidth}
        \includegraphics[trim = 0mm 0mm 0mm 0mm, clip, width=\textwidth]{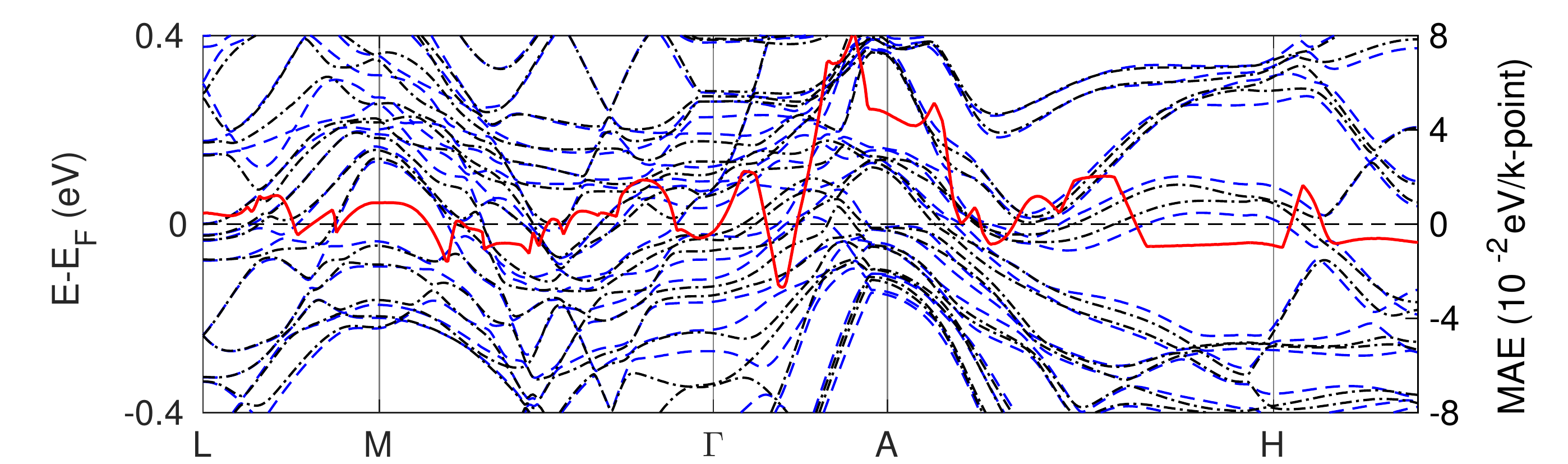}
        \caption{Band structure including SOC with magnetisation along 100 (black dash-dotted line) or 001-direction (blue dashed line) as well as the MAE contribution per k-point (red solid line). }
        \label{fig:Ta_MAEbands}
    \end{subfigure}	
	\caption{Atomic type and spin resolved band structure of Fe$_2$Ta with the colors red, green and blue indicating the contribution of $m=0$ (d$_{z^2}$), $m=1$ (d$_{xz}$ or d$_{yz}$) and $m=2$ (d$_{xy}$ or d$_{x^2 - y^2}$) states respectively, in (a)-(f). Black bands mean that the d-orbitals of given atomic type do not contribute significantly to the band in that region. (g) shows bands with SOC as well as $\mathbf{k}$-point resolved MAE contributions obtained via the magnetic force theorem.}
	\label{orbitalbands_Ta}
\end{figure*}

From the discussion above it is motivated to perform a careful analysis of the electronic band structure near the Fermi energy to obtain a better understanding of the MAE. Fig.~\ref{orbitalbands_Ta}a)-f) shows the spin polarized band structure through various high symmetry points in the Brillouin zone, without SOC, for Fe$_2$Ta, with spin up states on the left side and spin down states on the right side. Color coding is used to show the orbital character of the bands with red, green and blue indicating $m=0$ (d$_{z^2}$), $m=1$ (d$_{xz}$ or d$_{yz}$) and $m=2$ (d$_{xy}$ or d$_{x^2 - y^2}$) character, respectively, for different atomic types in the different rows. A black region on a band indicates that the given atomic type is not significantly contributing to the band in that region. The large number of bands present, even within one electronvolt from the Fermi surface, and complicated band structure with further complication due to hybridisation, makes analysis of the MAE in terms of the band structure difficult. Some observation can, nevertheless, directly be made. The $\Gamma$ point is often of particular importance since it has the highest symmetry. Here there are occupied and unoccupied spin up states very near the Fermi energy at this point, potentially allowing very strong effect from the SOC, especially since these states both show strong Ta contributions and Ta has the largest SOC constant. However, the unoccupied band is largely of $m=0$ character, while the occupied one is of $m=1$ character. Such states do not couple via SOC (see Table~\ref{SOmatrix}), whereby the potentially strong MAE contribution at $\Gamma$ is absent. 

To obtain information about which regions in reciprocal space are particularly important to the MAE, the band structures after applying SOC with magnetization along either 100 or 001 directions are plotted in Fig.~\ref{orbitalbands_Ta}g). From these bands the MAE contribution per $\mathbf{k}$-point can be evaluated using the magnetic force theorem\cite{Daalderop1990}, by taking the difference of the sum over occupied energy eigenvalues for different magnetization directions, which is also plotted (red line, right $y$-axis) in Fig.~\ref{orbitalbands_Ta}g). Since the MAE is positive in Fe$_2$Ta, regions with positive MAE contributions are expected to outweigh the negative regions. In agreement with the observation mentioned about $\Gamma$ above, there is a rather weak MAE contribution from the region around that point. Instead, it is clear that the most important region is that around the $A$-point where a large and positive MAE contribution is seen, while other regions show smaller values of varying sign, which one might expect to nearly cancel out in a Brillouin zone integration. From a first look at the bands in Fig.~\ref{orbitalbands_Ta}a)-f), the most important bands for the MAE at $A$ should be the highest occupied and lowest unoccupied ones, which are in both cases spin down with four-fold degeneracy. However, in Fig.~\ref{orbitalbands_Ta}g) one can identify the strongest positive MAE contribution where occupied 001-bands (blue dashed line) are shifted well below the corresponding 100-bands (black dash-dotted line). This occurs mainly for the highest occupied (also four-fold degenerate) spin up bands at $A$, whereby these should also be considered. The three sets of band which thus far appear most important at $A$ all have significant contributions from several atomic types and orbitals, in particular Ta and Fe$_1$, $m=1$ and $m=2$ states, but for the lowest unoccupied spin down bands, Fe$_2$ $m=1$ and $m=2$ are also important. This means that detailed analysis of the MAE contribution from the $A$-point is complicated since a large number of terms from Eq.~\ref{HybridisationSOC} must be considered. It is clear, however, that there is a significant Fe-Ta hybridisation in the relevant region and as was pointed out above, this allows for significant additions to the MAE. 

\begin{figure*}[hbt!]
	\centering
    \begin{subfigure}[b]{0.49\textwidth}
        \includegraphics[trim = 0mm 6.5mm 0mm 3mm, clip, width=\textwidth]{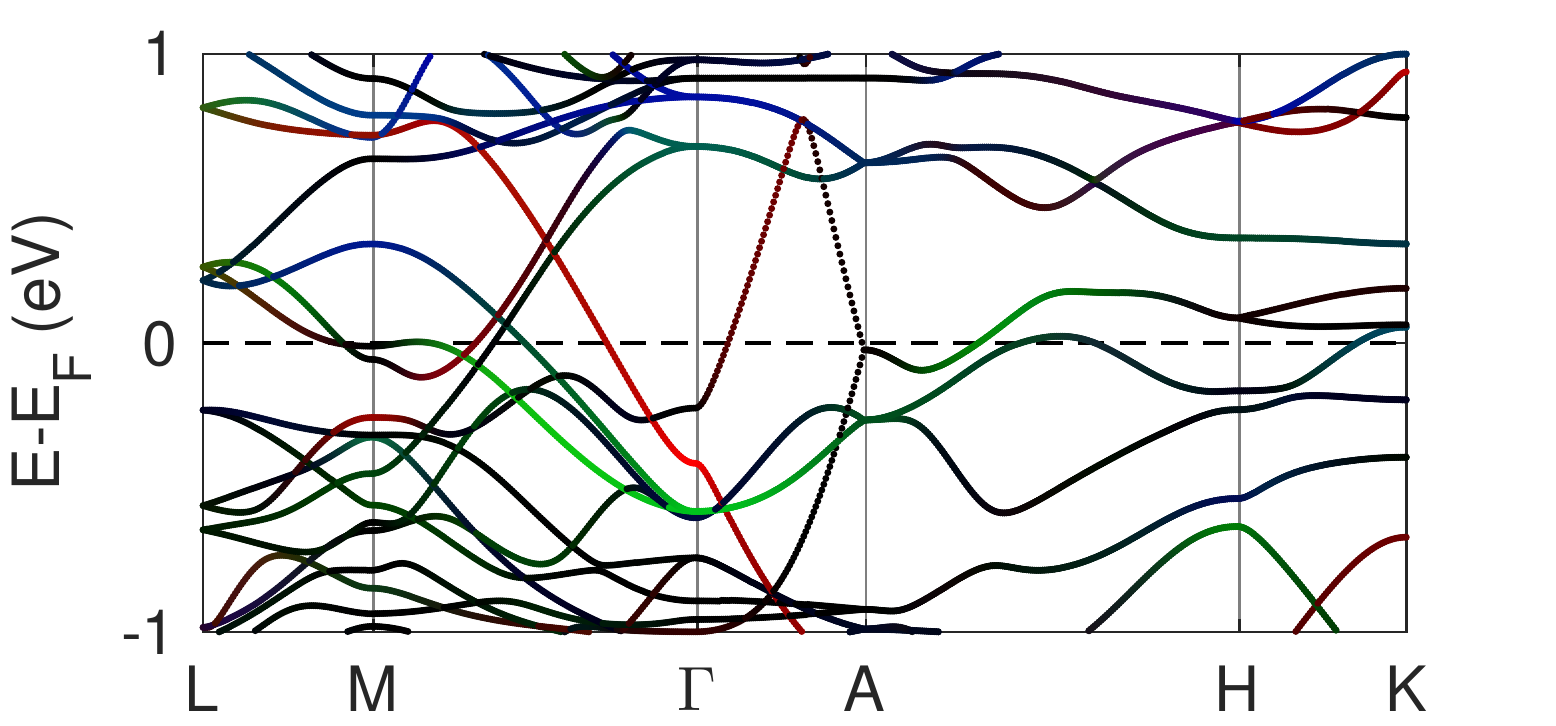}
        \caption{Fe$_1$, spin up. }
        \label{fig:W_Fe1_up}
    \end{subfigure}	
    \begin{subfigure}[b]{0.49\textwidth}
        \includegraphics[trim = 0mm 6.5mm 0mm 3mm, clip, width=\textwidth]{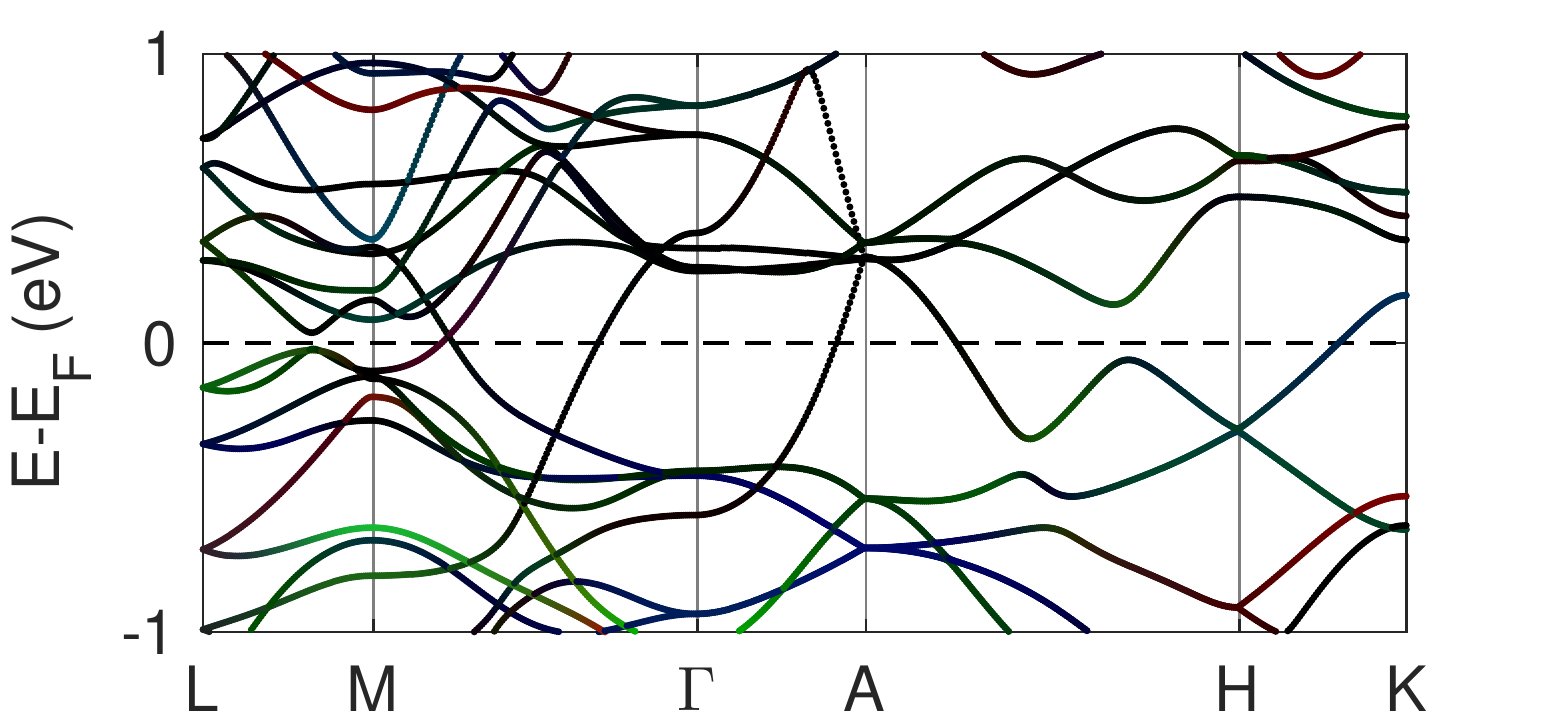}
        \caption{Fe$_1$, spin down. }
        \label{fig:W_Fe1_dn}
    \end{subfigure}	
    \begin{subfigure}[b]{0.49\textwidth}
        \includegraphics[trim = 0mm 6.5mm 0mm 3mm, clip, width=\textwidth]{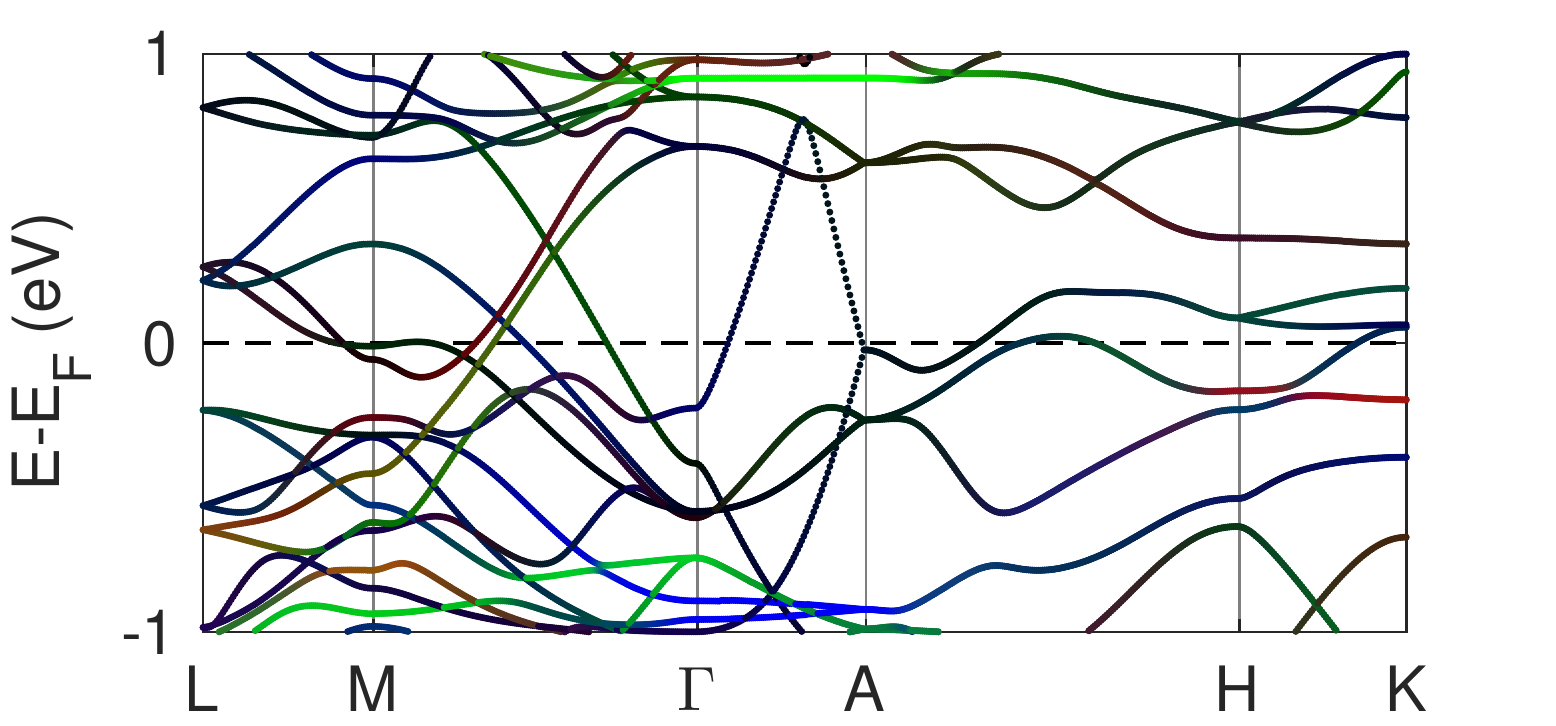}
        \caption{Fe$_2$, spin up. }
        \label{fig:W_Fe2_up}
    \end{subfigure}	
    \begin{subfigure}[b]{0.49\textwidth}
        \includegraphics[trim = 0mm 6.5mm 0mm 3mm, clip, width=\textwidth]{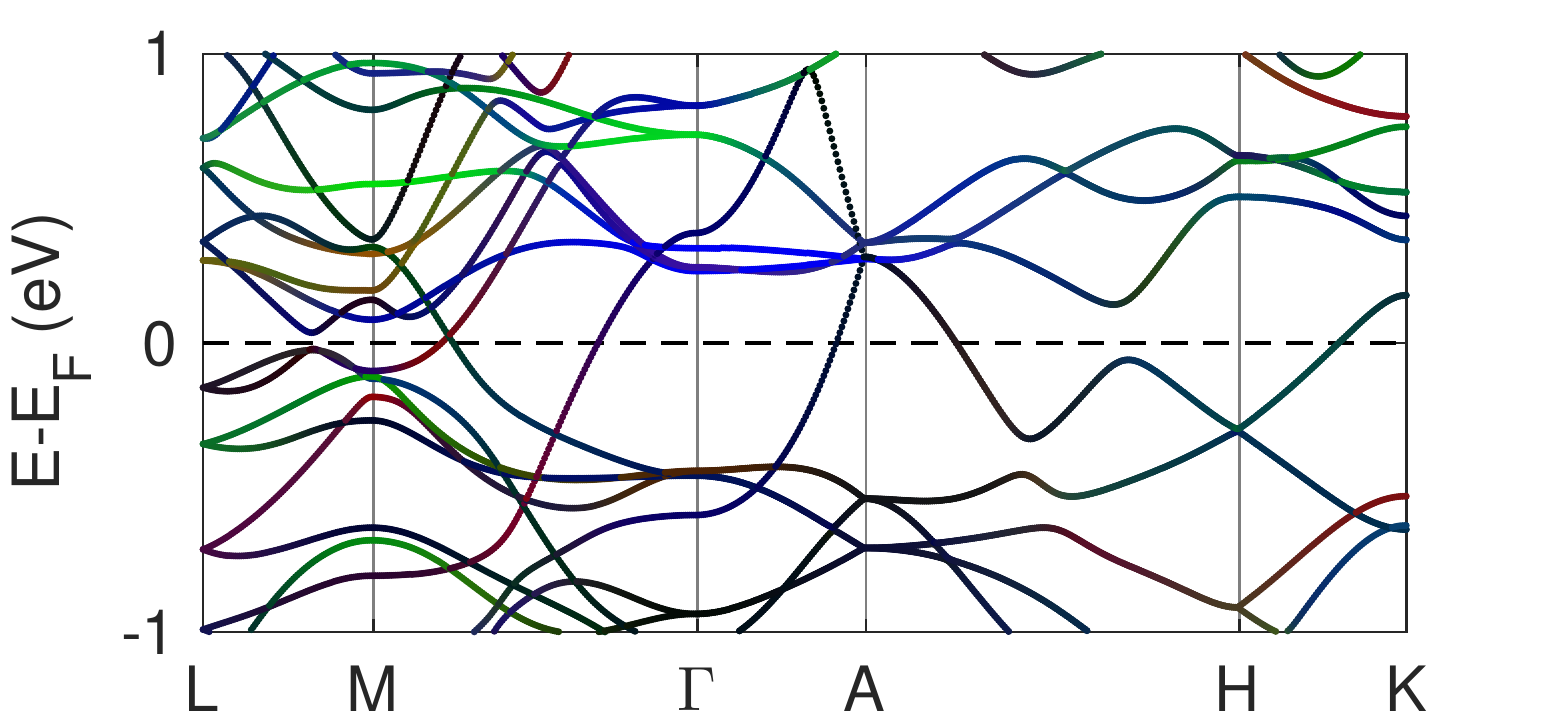}
        \caption{Fe$_2$, spin down. }
        \label{fig:W_Fe2_dn}
    \end{subfigure}	
    \begin{subfigure}[b]{0.49\textwidth}
        \includegraphics[trim = 0mm 1mm 0mm 3mm, clip, width=\textwidth]{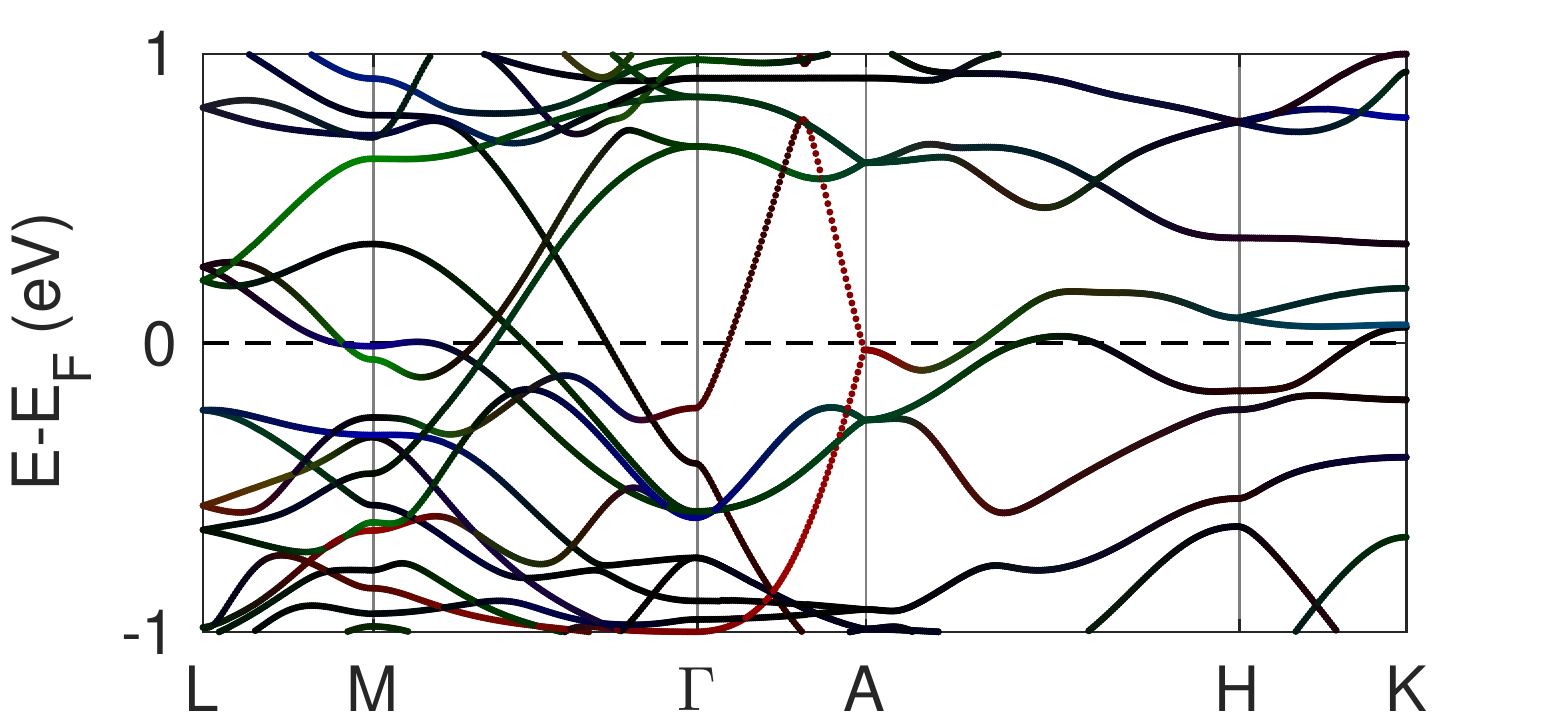}
        \caption{W, spin up. }
        \label{fig:W_up}
    \end{subfigure}	
    \begin{subfigure}[b]{0.49\textwidth}
        \includegraphics[trim = 0mm 1mm 0mm 3mm, clip, width=\textwidth]{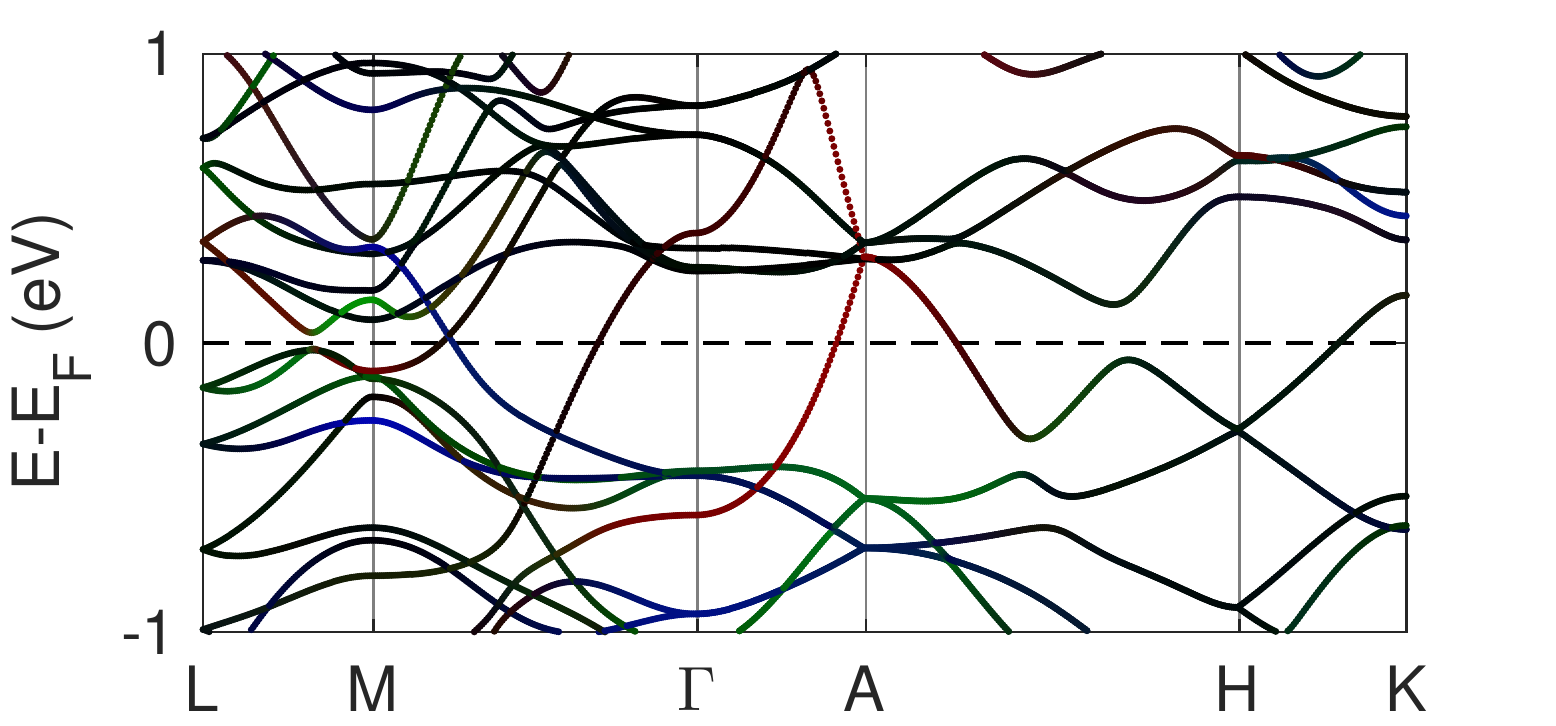}
        \caption{W, spin down. }
        \label{fig:W_dn}
    \end{subfigure}	
    \begin{subfigure}[b]{0.92\textwidth}
        \includegraphics[trim = 0mm 0mm 0mm 0mm, clip, width=\textwidth]{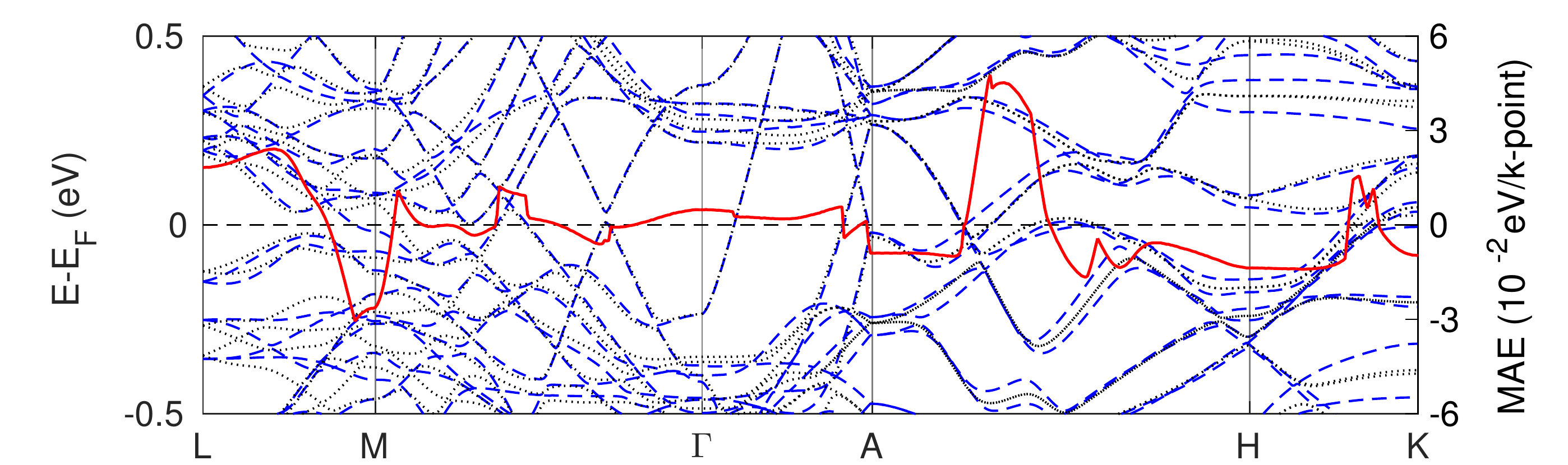}
        \caption{Band structure including SOC with magnetisation along 100 (black dash-dotted line) or 001-direction (blue dashed line) as well as the MAE contribution per k-point (red solid line). }
        \label{fig:W_MAEbands}
    \end{subfigure}	
	\caption{Atomic type and spin resolved band structure of Fe$_2$W with the colors red, green and blue indicating the contribution of $m=0$ (d$_{z^2}$), $m=1$ (d$_{xz}$ or d$_{yz}$) and $m=2$ (d$_{xy}$ or d$_{x^2 - y^2}$) states respectively, in (a)-(f). Black bands means that the d-orbitals of given atomic type do not contribute significantly to the band in that region. (g) shows bands with SOC as well as $\mathbf{k}$-point resolved MAE contributions obtained via the magnetic force theorem.}
	\label{orbitalbands}
\end{figure*}

Fig.~\ref{orbitalbands} contains the same type of information as Fig.~\ref{orbitalbands_Ta}, but for Fe$_2$W. Since Fe$_2$W also has a uniaxial (positive) MAE, positive regions are expected to dominate the MAE contributions in Fig.~\ref{orbitalbands}g). In similarity with the Fe$_2$Ta case, there are large regions of small contributions with varying sign, which one would expect to nearly vanish in an integration. In particular, the important $\Gamma$-point provides a weak contribution, which can be understood from the relatively large separation in energy between the highest occupied and lowest unoccupied states, compared to other regions. The most important positive contributions to the MAE stem from the $L$-neighbourhood, as well as a region along the path $A-H$, while there is a significant negative region around $M$, which might partially explain why the MAE of Fe$_2$W is weak. In the important region along the $A-H$ path, there are two spin up bands nearly parallel to each other. These are on opposite sides of the Fermi energy where the $\mathbf{k}$-point resolved MAE is strongest, and can therefore contribute to the MAE. Both bands are mainly of W and Fe$_1$ $m=1$ character. From the SOC matrix elements in the appendix, one finds that states of same spin and $m$ value yield a positive (uniaxial) contribution to the MAE. Furthermore, the Fe-W hybridisation allows the large W SOC to make the coupling strength large and this explains the large positive MAE coming from that part of the $A-H$ path. 

At the $L$-point, a significant positive source of MAE is found in the highest occupied spin up states which are mainly Ta and Fe$_1$ $m=1$, since the 001 bands are shifted below the 100 bands. This situation is reversed as one moves along the $L-M$ path and the change of sign in the $\mathbf{k}$-resolved MAE appears to coincide with the spin down bands which are unoccupied at $L$ becoming occupied near $M$. The presence of many bands with significant hybridisation effects makes it difficult to pinpoint states coupling via SOC which are particularly important to the MAE along the $L-M$ path. Nevertheless, it should be pointed out that once again there is significant Fe-W hybridisation, so that the strong W SOC can increase the MAE. Since there is a limited 5d contribution to the DOS at the Fermi energy, there can only be significant 3d-5d hybridisation near the Fermi energy in a limited region of the Brillouin zone. Nevertheless, the reciprocal space analysis of the electronic structure and MAE contributions reveals that the MAE is mainly determined by those regions in the Brillouin zone where there is notable 3d-5d hybridisation, in both Fe$_2$Ta and Fe$_2$W.

\begin{figure}[hbt!]
	\centering
	\includegraphics[width=0.49\textwidth]{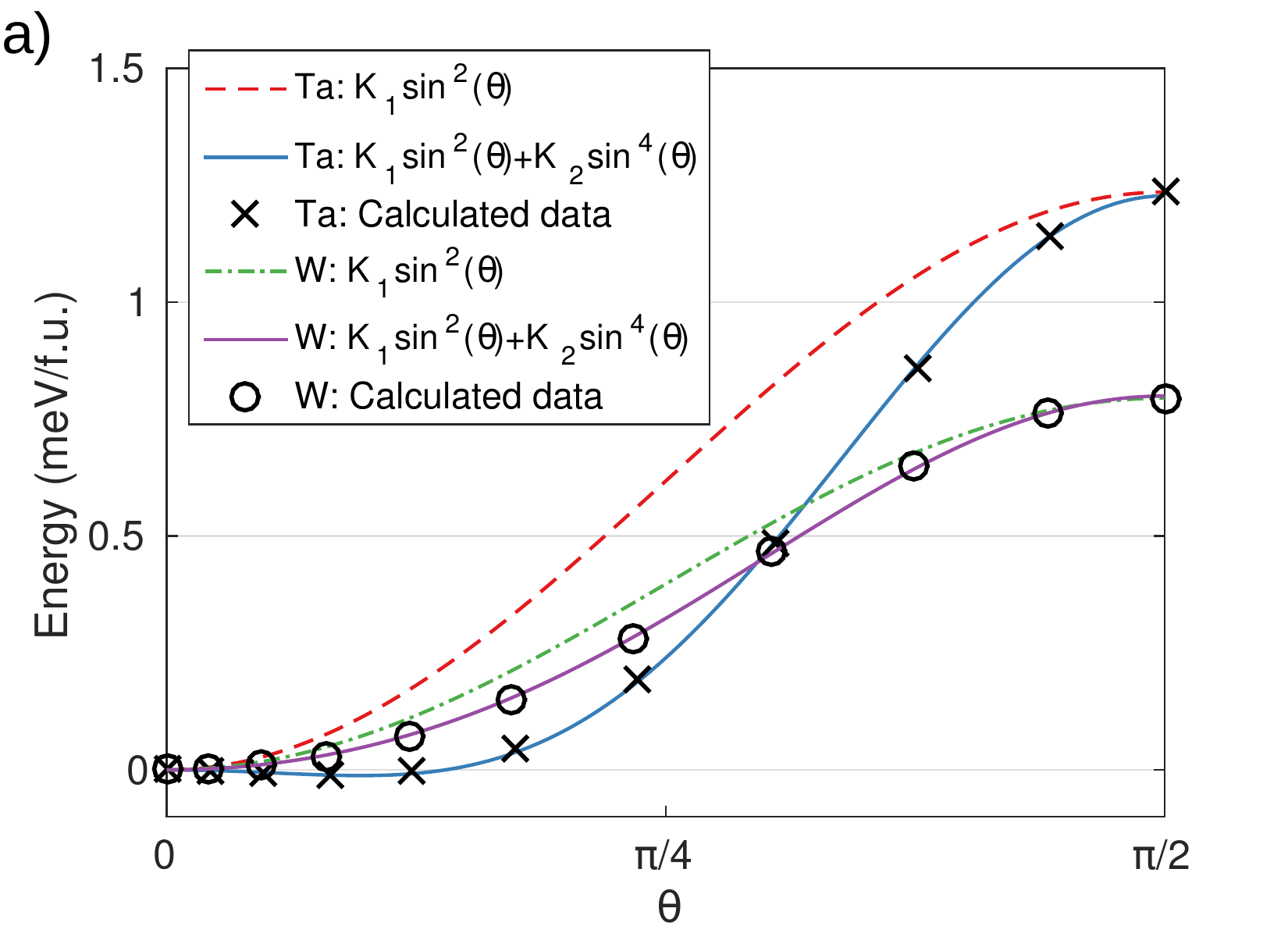} 
	\includegraphics[width=0.49\textwidth]{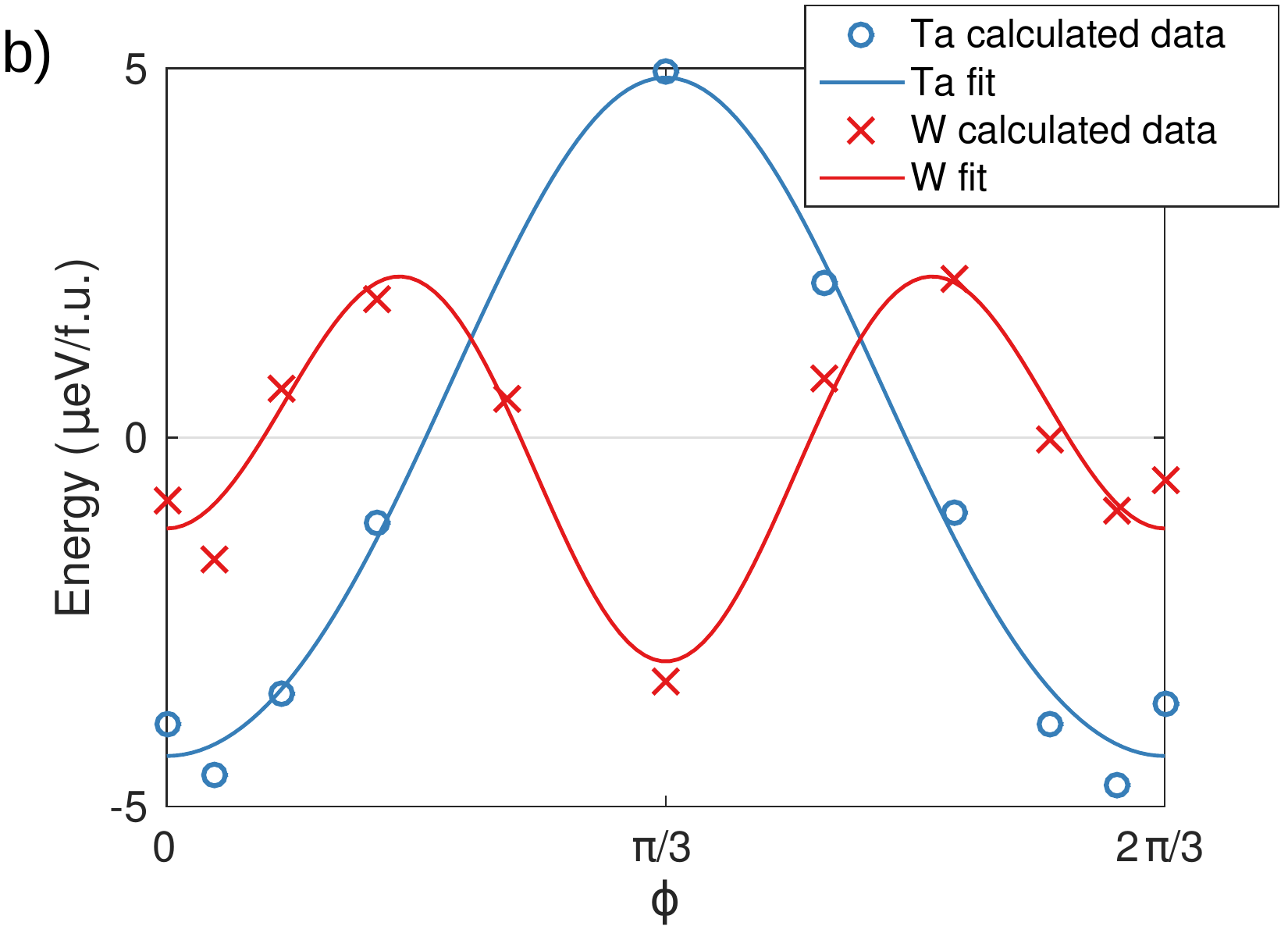} 
	\caption{Energy as a function of the polar angle $\theta$ between the $c$-axis and the magnetization direction in a) and Energy as a function of the azimuthal angle $\phi$ with $\theta=\frac{\pi}{2}$ in b). The fit in b) is to a function $E(\theta=\frac{\pi}{2}, \phi)=C_1 + C_2 \cos 3\phi + C_3 \cos 6\phi$ and $C_2 \cos 3\phi + C_3 \cos 6\phi$ is plotted. }
	\label{fig.Eoftheta}
\end{figure}

As both quantities are due to the SOC, Bruno\cite{Bruno1989} pointed out the close relation between magnetocrystalline anisotropy and orbital moments and showed, using perturbation theory on a tight binding model, that if deformations of the Fermi surface can be neglected and the MAE is dominated by spin-diagonal coupling, the MAE and orbital magnetic moment anisotropy are proportional. If coupling between minority spin states dominates the SOC, a maximum orbital magnetic moment is expected in the easy direction of magnetization, as is seen in the case of Fe$_2$Ta in Table~\ref{table2}. If, on the other hand, the SOC is dominated by the coupling between majority spin states, a maximum orbital magnetic moment is expected along the hard magnetisation axis, as is seen in the case of Fe$_2$W. This is consistent with the observation made in Fig.~\ref{dos}, that the Fe$_2$Ta DOS($E_\text{F}$) is dominated by minority spin states, while the opposite is true for Fe$_2$W. For a further analysis of the relation between MAE and $m_L$ in the studied systems, energy and orbital moments have been computed as functions of the angle $\theta$ (with $\phi=0$) when the magnetization is along $\hat{\mathbf{n}} = (\sin \theta \cos \phi, \sin \theta \sin \phi , \cos \theta )$. The result for the energy as function of $\theta$ is shown in Fig.~\ref{fig.Eoftheta}. The second order perturbation theory for a uniaxial system leads to the conclusion that the energy as function of $\theta$ follows the relation
\begin{equation} \label{2ndorderKs}
E(\theta) = K_0 + K_1 \sin^2 \theta , 
\end{equation}
with isotropic energy $K_0$. This is merely the first part of the longer expansion
\begin{align}
E(\theta, \phi) & = K_0 + K_1 \sin^2 \theta + K_2 \sin^4 \theta + \nonumber \\ 
				 & + K_3 \sin^6 \theta \left( 1 + k_{3,3} \cos 3\phi + k_{3,6} \cos 6\phi \right) + ...
\end{align}
valid for a uniaxial crystal with three-fold rotational symmetry about the $z$-axis, such as the one studied here. For a system where the MAE is well described by second order perturbation theory, one expects that the energy is well fitted by Eq.~\ref{2ndorderKs} and that $K_i$ is vanishingly small for $i>1$. As seen in Fig.~\ref{fig.Eoftheta}a), fitting the energy as function of angle between magnetization direction and 001-direction to $K_1 \sin^2 \theta$ provides an unsatisfactory curve for $E(\theta)$ for both Fe$_2$Ta and Fe$_2$W, while including also the term $K_2 \sin^4 \theta$ yields an excellent fit (for the fit to $K_1 \sin^2 \theta$, $K_1$ was simply set to $E(\pi/2)-E(0)$, while the fit to $K_1 \sin^2 \theta + K_2 \sin^4 \theta$ was done with the method of least squares). This indicates that second order perturbation theory provides a quantitatively inaccurate description of the MAE in the studied compounds, while fourth order terms should provide an accurate description with higher (than fourth) order corrections being small. Clearly, the fit to $K_1 \sin^2 \theta$ is significantly better in the case of Fe$_2$W than for Fe$_2$Ta. This indicates that restriction to second order perturbation theory, rather than fourth, is a better approximation for the W compound, which might be related to the smaller contribution of the 5d atom to the DOS($E_\text{F}$), making the assumption of a small $\xi$ more realistic. 

The anisotropy constants obtained from the fitting to $K_1 \sin^2 \theta + K_2 \sin^4 \theta$ are listed in Table~\ref{table3}. As was already anticipated from Fig.~\ref{fig.Eoftheta}a), $K_2$ is of more importance in Fe$_2$Ta and in fact it is of opposite sign and significantly bigger than $K_1$. In the case where $K_1$ and $K_2$ have the same sign, the $\theta$-derivative of $E(\theta) = K_1 \sin^2 \theta + K_2 \sin^4 \theta$ has only two zeros for real $K_i$, namely $\theta=0$ and $\theta=\pi/2$, whereby the easy and hard magnetization directions will occur at these angles. For opposite signs of $K_1$ and $K_2$, an additional zero occurs at 
\begin{equation}
\theta = \sin^{-1}(\sqrt{-\frac{K_1}{2 K_2}})
\end{equation}
and for Fe$_2$Ta there is a minimum in the energy at approximately $\theta = 0.15 = 8.8^\circ$. The easy magnetization direction is thus expected at this angle rather than at $\theta=0$, so the material strictly speaking does not have a uniaxial magnetization. For Fe$_2$W, both constants are positive so $\theta=0$ is the easy axis. Although in this case the magnitude of $K_1$ is greater than $K_2$, the latter is not negligible. 
\begin{table}
\caption{\label{table3} Anisotropy constants $K_1$,  $K_2$ and $\tilde{K}_3=K_3(1+k_{3,3} + k_{3,6})$ from least squares fitting of $E(\theta,\phi=0)$ to $K_1 \sin^2 \theta + K_2 \sin^4 \theta$ or $K_1 \sin^2 \theta + K_2 \sin^4 \theta + \tilde{K}_3 \sin^6 \theta$ (see Fig.~\ref{fig.Eoftheta}).}
\begin{ruledtabular}
\begin{tabular}{l l l l}
 				& $K_1~(\text{meV/f.u.})$ & $K_2~(\text{meV/f.u.})$ & $\tilde{K}_3~(\text{meV/f.u.})$ \\
\hline  
Fe$_2$Ta 		& -0.27	& 1.50  &				\\
Fe$_2$Ta 		& -0.19	& 1.23  & 0.19  	\\ \hline
Fe$_2$W			& 0.50		& 0.30  &				\\
Fe$_2$W			& 0.45		& 0.46	 &	-0.11 	\\
\end{tabular}
\end{ruledtabular}
\end{table}

Table~\ref{table3} also contains parameters from a fit to $K_1 \sin^2 \theta + K_2 \sin^4 \theta + \tilde{K}_3 \sin^6 \theta$. This indicates non-negligible values of $\tilde{K}_3$ for both compounds and, in the case of Fe$_2$Ta, it is of the same magnitude as $K_1$. However, it is not clear how many fitting parameters are reasonable to include with the given numerical accuracy. Comparison to a fit from a calculation with only $2\times 10^4$ $\mathbf{k}$-points yields a value smaller by a factor of one third for Fe$_2$Ta, indicating that the numerical accuracy might be insufficient. However, more accurate calculations become prohibitively computationally demanding.  

Typically, in uniaxial systems which do not possess strong SOC, the variation in energy for rotations of the magnetisation direction in the plane is small. This makes it challenging and computationally heavy to compute the in plane magnetic anisotropy (this might differ in, for example, actinide systems, where even cubic materials can have enormous MAE~\cite{Lander1990}). Nevertheless, the energy as a function of $\phi$ with $\theta=\frac{\pi}{2}$ was computed and the result is shown in Fig.~\ref{fig.Eoftheta}b). The calculated points have been fitted to $E(\theta=\frac{\pi}{2}, \phi)=C_1 + C_2 \cos 3\phi + C_3 \cos 6\phi$ ($C_2$ and $C_3$ should correspond to $K_3 k_{3,3}$ and $K_3 k_{3,6}$, respectively) and $C_1$ has been subtracted from the calculated points and fitted curves. As expected, the variations in Fig.~\ref{fig.Eoftheta}b) are much smaller, by nearly three orders of magnitude, than the variations seen in Fig.~\ref{fig.Eoftheta}a). It is difficult to say whether the deviations between the computed points and the fitted lines are mainly due to limitations in the numerical accuracy or because of neglecting higher order terms. 

\begin{figure}[hbt!]
	\centering
	\begin{subfigure}[b]{0.49\textwidth}
        \includegraphics[width=\textwidth]{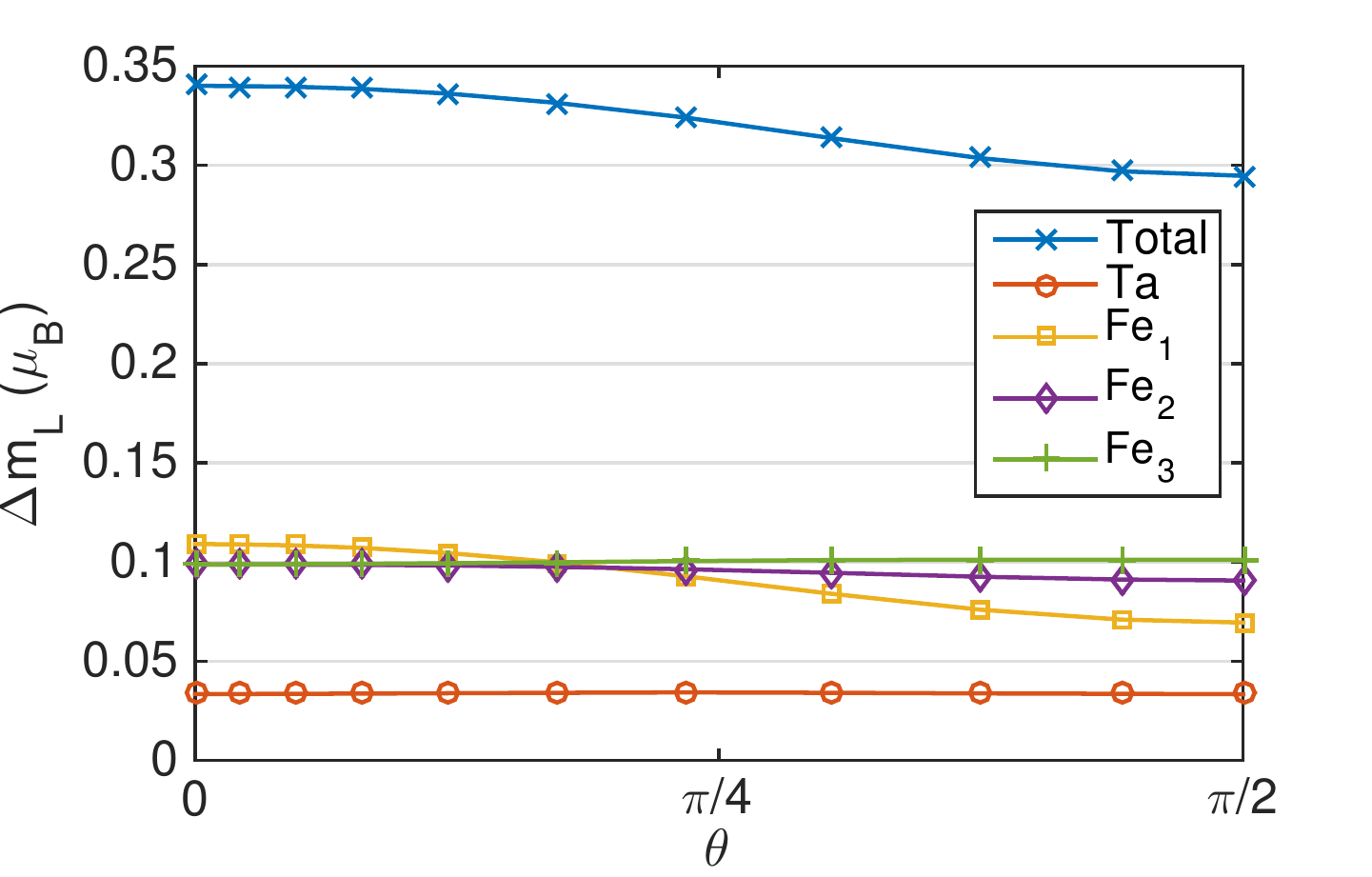}
        \caption{Orbital magnetic moment as function of $\theta$ for Fe$_2$Ta.}
        \label{fig:MAEofTheta}
    \end{subfigure}
    	\begin{subfigure}[b]{0.49\textwidth}
        \includegraphics[width=\textwidth]{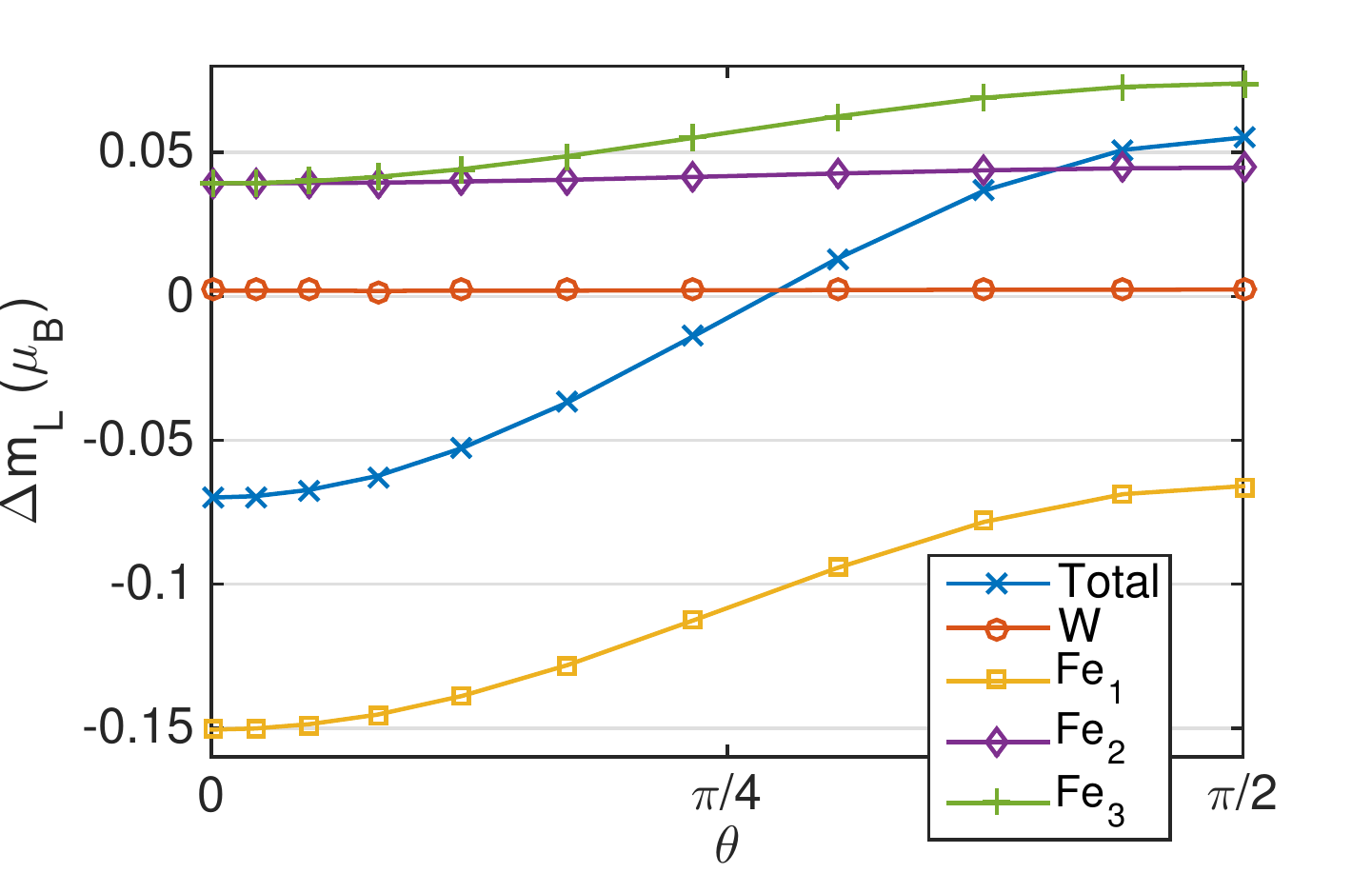}
        \caption{Orbital magnetic moment as function of $\theta$ for Fe$_2$W.}
        \label{fig:LofTheta}
    \end{subfigure}	
    	\begin{subfigure}[b]{0.49\textwidth}
        \includegraphics[width=\textwidth]{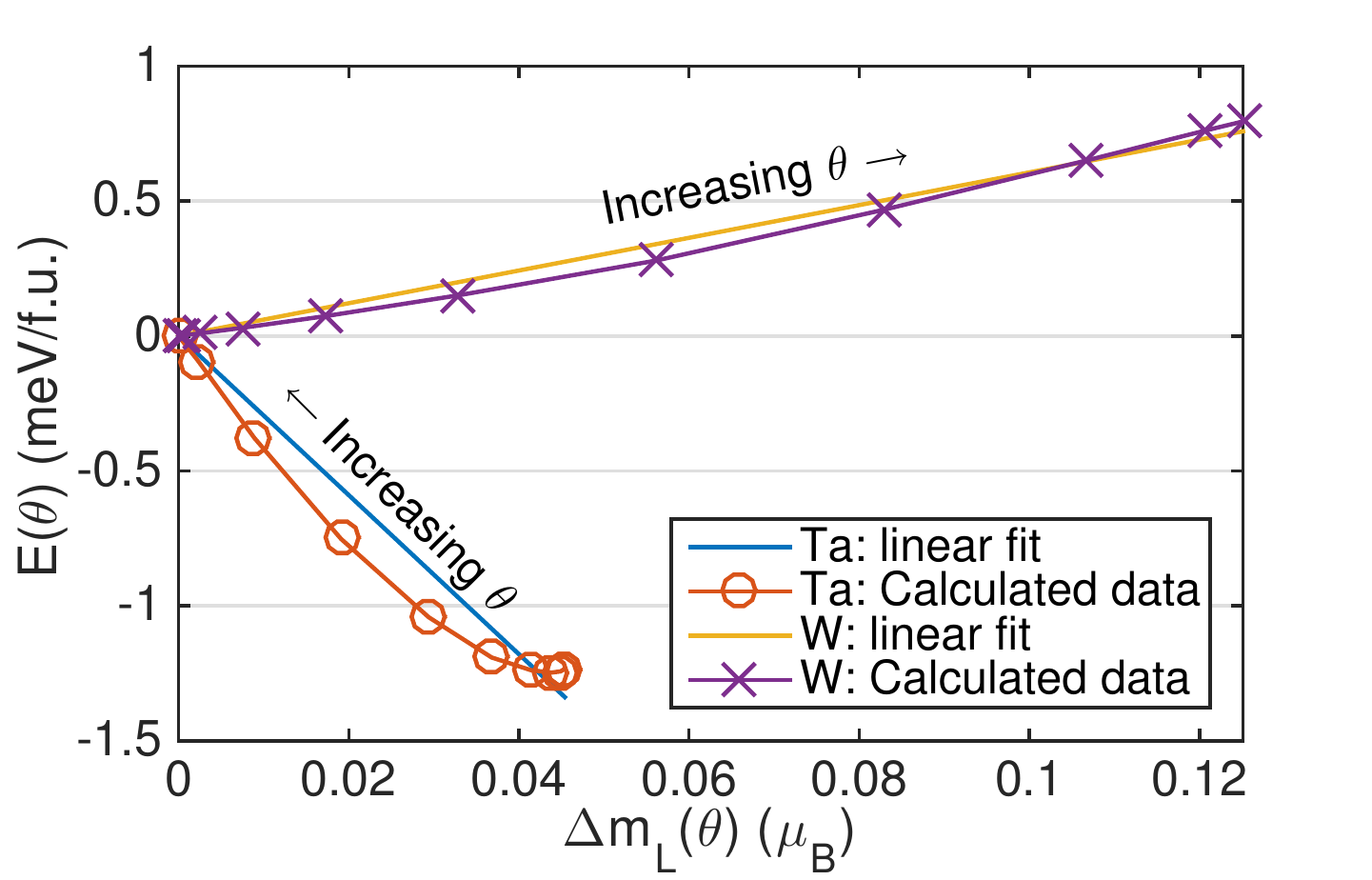}
        \caption{Change in energy versus change in orbital moment as $\theta$ is varied from $0$ to $\pi/2$.}
        \label{fig:EofL}
    \end{subfigure}		
	\caption{Energy and orbital magnetic moments as function of the angle $\theta$ between then magnetization direction and the 001 axis. }
	\label{EandLofTheta}
\end{figure}

Fig.~\ref{EandLofTheta} shows how the orbital magnetic moments vary with magnetization direction for Fe$_2$Ta (a) and Fe$_2$W (b).  In both materials the greatest contribution to the orbital moment anisotropy is due to the Fe$_1$ atom. The Fe$_2$ and Fe$_3$ atoms have identical orbital magnetic moments at $\theta=0$, as expected from symmetry, while they deviate from one another at other directions. The compounds differ in the sign of the variation of the orbital magnetic moment with $\theta$, although they both have same sign of $K_1+K_2$. In Fig.~\ref{EandLofTheta}c), which shows a plot of the energy as function of $\theta$ vs the anisotropy in total orbital magnetic moment as function of $\theta$, this appears as a difference in the sign of the slope of the curves. As was previously mentioned, this can be understood in terms of the DOS($E_\text{F}$) which is mainly due to the majority spin channel in Fe$_2$W and mainly due to the minority spin channel for Fe$_2$Ta. According to the work of Bruno\cite{Bruno1989}, this should lead to approximate proportionality between $\Delta m_L (\theta)$ and $E(\theta)$, but with opposite signs in the proportionality constants. However, that was based on second order perturbation theory, and as was seen above, fourth order perturbation theory is expected to be necessary for a quantitative description of the magnetic anisotropy in these materials, especially in Fe$_2$Ta. Fig.~\ref{EandLofTheta}c) also shows a linear fit to the curves for $E(\theta)$ vs $\Delta m_L (\theta)$. For Fe$_2$W, the linear fit provides a reasonable description of the curve, while in Fe$_2$Ta the deviation from linearity is more pronounced. This might largely be because of strong spin polarisation of the DOS at $E_\text{F}$ for Fe$_2$W, which makes the approximation that only spin diagonal SOC contributes to the magnetic anisotropy more reasonable. Although the DOS($E_\text{F}$) in Fe$_2$Ta is dominated by minority spin states, the contribution from the majority spin channel is significant, whereby neglecting spin-off diagonal contributions is questionable. Furthermore, the stronger contribution from the 5d states could also affect the relation between MAE and orbital moment anisotropy in that direction, consistent with previous observations\cite{Andersson2007} of non-proportionality between orbital magnetic moment and anisotropy in energy systems with significant 3d-5d hybridisation.

As the MAE depends sensitively on the band structure around the Fermi energy, it can be controlled by tuning the band structure around the Fermi energy. In practice this can be done, for example, by alloying, which will be explored next by considering the alloy Fe$_2$Ta$_{1-x}$W$_x$. Due to the complicated electronic structure, which was illustrated in Fig.~\ref{fig:Ta_MAEbands}) and Fig.~\ref{fig:W_MAEbands}), it is difficult to predict the effect of alloying on properties such as the MAE without explicitly doing calculations to evaluate the properties. For the system studied here it is also of interest to investigate where the transition from ferro- to ferrimagnetism occurs. The virtual crystal approximation\cite{Faulkner19821} (VCA), in which the alloyed atoms are exchanged for virtual atoms with non-integer effective atomic numbers, $Z$, which on average have the right ionic charge and number of electrons for a given alloy concentration, will be used to treat the disorder. The VCA, although simple compared to more sophisticated single site approximations, such as the coherent potential approximation (CPA), often provides a good average description for properties such as magnetic moments\cite{Victora1984,Delczeg-Czirjak2014,Edstrom2015,Niarchos2015}, especially for neighbours in the periodic table and small alloy concentrations\cite{Faulkner19821}. For delicate properties, like the MAE, on the other hand, the VCA has often been seen to result in quantitative overestimates compared to CPA calculations\cite{Turek2012, Edstrom2015}, super cell calculations\cite{Neise2011,Steiner2016} or experiments\cite{Andersson2006,Reichel2014,Edstrom2015}. Nevertheless, one should still be able to observe correct qualitative trends in the MAE from the VCA and it will be applied also for this property.

\begin{figure}[hbt!]
	\centering
        \includegraphics[width=0.48\textwidth]{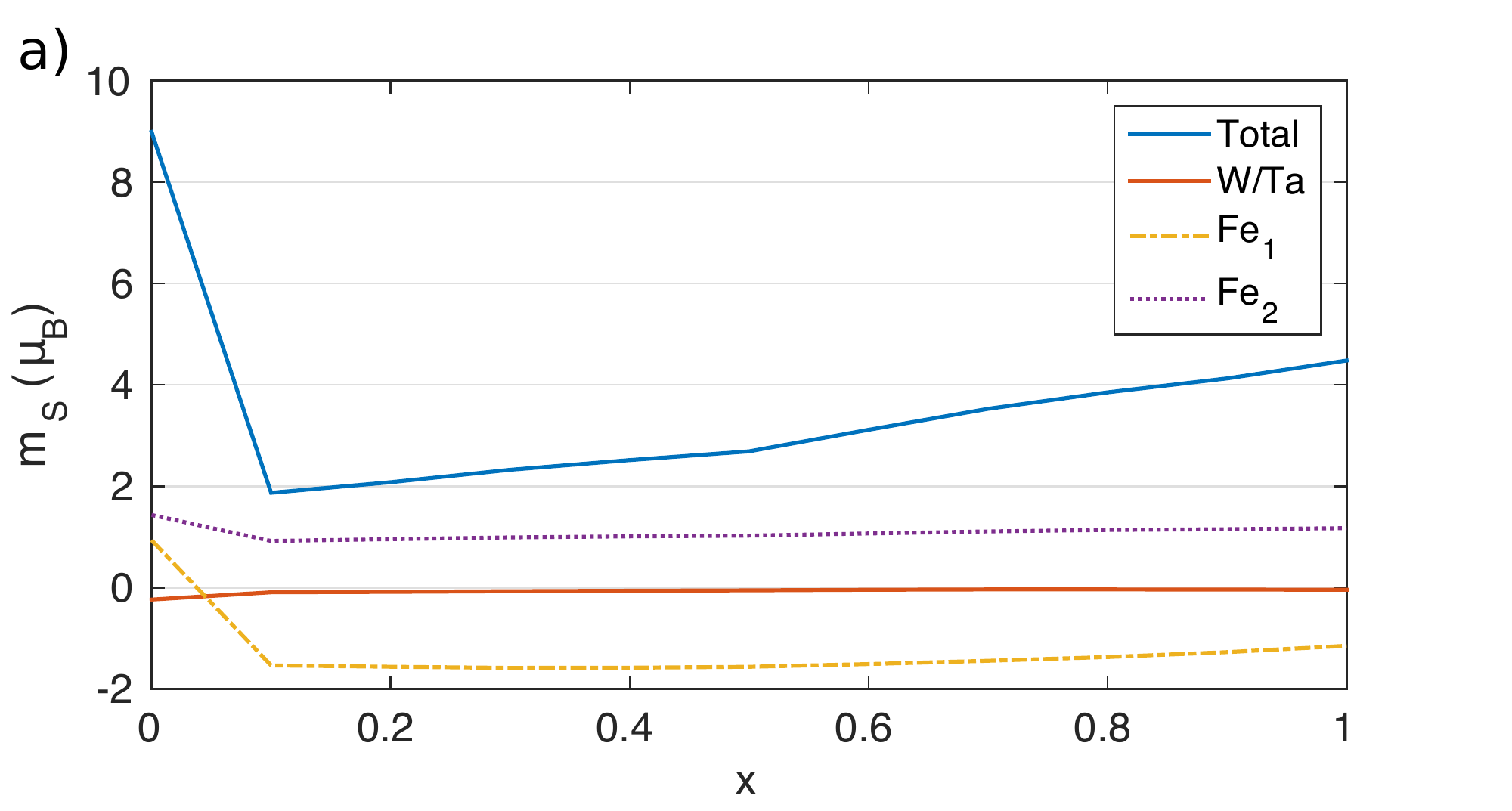}
        \includegraphics[width=0.48\textwidth]{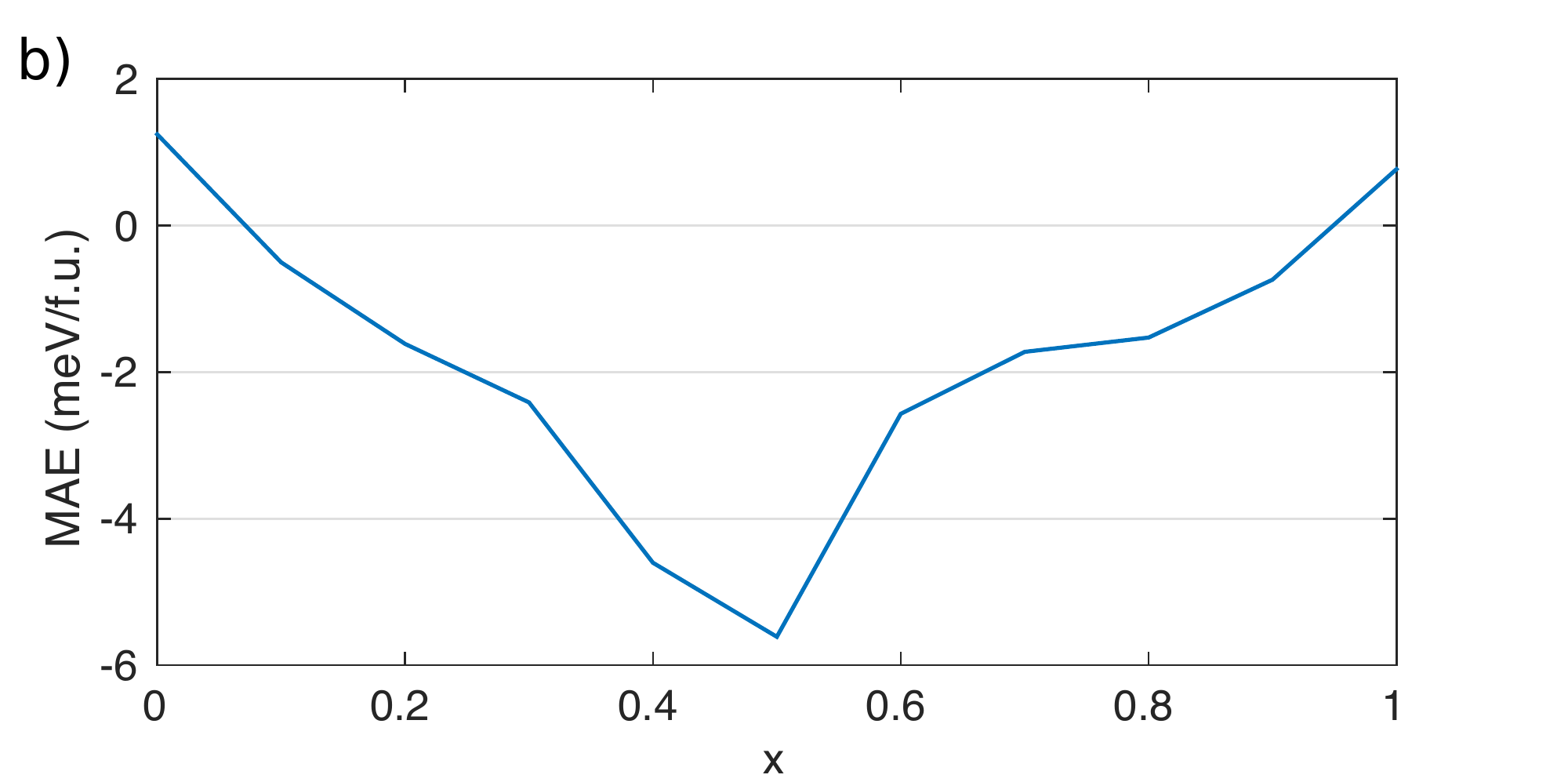}
	\caption{a) Spin magnetic moments and b) MAE (computed as total energy differences for magnetizations along 100 and 001 directions) as functions of $x$ in Fe$_2$Ta$_{1-x}$W$_x$.}
	\label{ms_MAE_alloy}
\end{figure}

Calculations were performed for values of $x$ in increments of 0.1. A calculation for $x=0.1$ revealed that this is enough for the magnetic ordering to transition into the ferrimagnetic ordering observed also for Fe$_2$W. A complete structural relaxation, using spin polarized calculations neglecting SOC, was thus performed for $x=0.1$. The resulting lattice parameters are $a=4.771~\text{\AA}$ and $c=7.847~\text{\AA}$. Lattice parameters for $0.2 \leq x \leq 0.9$ were calculated by linear interpolation between the values obtained for $x=0.1$ and $x=1.0$. Calculations including SOC were then performed for the whole range of alloys and the resulting spin magnetic moments (for magnetization along the $c$-axis) and MAEs are presented in Fig.~\ref{ms_MAE_alloy}. A large decrease in total spin magnetic moment is seen when going from $x=0$ to $x=0.1$, due to the change in sign of the Fe$_1$ spin moment, but also because of an accompanying reduction in size of the Fe$_2$ moment. For $x$ greater than 0.1, the total spin magnetic moment monotonically increases until $x=1.0$. This appears to be from a combination of decrease in size of the Fe$_1$ moment and increase in size of the Fe$_2$ moment. The MAE decreases with $x$ until it reaches a minimum at $x=0.5$ and then increases until $x=1.0$. Hence, the largest positive values of the MAE are obtained for the end compounds and it cannot be increased by the alloying considered here. A negative in-plane anisotropy of very large magnitude is seen for $x=0.5$. However, it is important to remember that the VCA is expected to overestimate the magnitude of the MAE, whereby the real value might be of smaller magnitude. 

A comprehensive computational study has been performed for the hexagonal Laves phase compounds Fe$_2$Ta and Fe$_2$W, with focus on the important intrinsic magnetic properties saturation magnetization and MAE. For Fe$_2$W, a new ferrimagnetic ground state has been suggested, different from that found in earlier computational work~\cite{Kumar2014}. In the case of Fe$_2$Ta, a similar magnetic ordering is found as in preceding calculations~\cite{Kumar2014}, but an opposite sign is found in the MAE. The discrepancies in comparison with earlier calculations calls for further experimental efforts to unambiguously determine the magnetic properties of these compounds. 

The MAE has been carefully analysed in terms of the electronic structure and by using the magnetic force theorem to compute $\mathbf{k}$-point resolved contributions to the MAE. Because the density of states at the Fermi energy is dominated by 3d states, 5d-states can only contribute notably to the MAE in small regions of the Brillouin zone. Nevertheless, it is found that the MAE originates mainly from regions in the Brillouin zone where there is a strong 3d-5d hybridisation, allowing the strong SOC of the 5d atoms to increase the MAE.

The main motivation to study uniaxial 3d-5d compounds is the possibility to have a very large MAE, such as the value of $6.6~\text{MJ/m}^3$~\cite{Coey2011} observed in FePt. When a significant amount of magnetic 3d elements is included, this can be combined with large saturation magnetisation and a high Curie temperature. Among the compounds studied here, the MAEs calculated are quite modest compared to that seen in FePt. In addition, for Fe$_2$W, a ferrimagnetic ordering is found, resulting in a low saturation magnetisation. Nevertheless, whether a material is useful for a given application depends on a combination of the mentioned intrinsic parameters. For example, in the context of permanent magnets, the hardness parameter 
\begin{equation}
\kappa=\sqrt{\frac{K}{\mu_0 M^2}},
\end{equation}
with MAE $K$ and saturation magnetisation $M$, can be used to determine whether a material has potential to exhibit a reasonable coercive field and be used as a permanent magnet\cite{Coey2011,Hirosawa2015}. $\kappa$ is required to be greater than unity but the microstructural engineering to obtain the desired properties of a permanent magnet should be easier with larger $\kappa$ and Hirosawa~\cite{Hirosawa2015} suggested $\kappa>1.4$ to be demanded from potential permanent magnet materials. For the materials studied here one finds $\kappa=1.8$ for Fe$_2$Ta and $\kappa=2.9$ for Fe$_2$W, from the data in Table~\ref{table2}, well above the requirement put forward by Hirosawa. These large values of $\kappa$ appear largely because of the modest saturation magnetisations and the energy product of a permanent magnet will be limited by this. In both materials the saturation magnetisation is below the value of $\mu_0 M_\text{s} = 1.6~\text{T}$~\cite{Coey2011} found in the powerful Nd$_2$Fe$_{14}$B magnet. However, at least in Fe$_2$Ta the saturation magnetisation is greater than $0.48~\text{T}$ seen in BaFe$_{12}$O$_{19}$ ferrite magnets, potentially making the compound technologically interesting as an intermediate alternative between rare-earth and ferrite magnets. 

Experimental work has reported a Curie temperature of 550 K in Fe$_2$W\cite{Koten2015}, which should be sufficient for many technological applications. As a useful extension of the current work, it would be interesting to compute the Curie temperatures of Fe$_2$W and Fe$_2$Ta, e.g., by calculating the Heisenberg exchange parameters from first principles and using these as input to the mean field approximation or Monte Carlo simulations. This would reveal whether Fe$_2$Ta also has a high enough Curie temperature to be technologically interesting and might also shed further light on the issue regarding the magnetic ordering of Fe$_2$W. 

To investigate the possibility of enhancing the relevant properties, alloying of W and Ta has been considered in calculations for Fe$_2$Ta$_{1-x}$W$_x$, with the disorder treated in the virtual crystal approximation. These calculations indicate that the transition from ferro- to ferrimagnetic ordering occurs for $x$ smaller than 0.1 and that the MAE is significantly reduced and mainly strongly negative in the alloy. For technological purposes this does not appear promising. However, there are various isostructural 3d-5d compounds, such as Mn$_2$Ta, Co$_2$Ta or Fe$_2$Hf~\cite{BLAZINA1989247,NISHIHARA198775} and one might also consider alloys among these. Allowing for 3d or 4d atoms to substitute the 5d atom gives further possibilities~\cite{NISHIHARA198775}. As a next step, it should be worthwhile to investigate ternary or quaternary phase diagrams for magnetic 3d elements combined with 5d and other elements in uniaxial crystals. Numerous such phases which have not been properly characterized in terms of magnetic properties should exist and the type of computational methods used in this work should be of great value in identifying interesting materials. 

I would like to thank Yaroslav Kvashnin for critically reading and providing useful comments on the manuscript. I'm also grateful to J\'{a}n Rusz and Olle Eriksson for discussions and for encouragement to pursue this work. Computational work has been performed with resources from the Swedish National Infrastructure for Computing (SNIC) at the National Supercomputer Centre (NSC) in Link\"oping. 

\appendix
\section{Matrix elements of the spin-orbit operator}\label{AppA}

If $\ket{i}$ is a single particle eigenstate to an unperturbed Hamiltonian with no SOC, the total shift in the energy $E_i$ due to $H_\text{SOC} = \xi \mathbf{\hat{l}} \cdot \mathbf{\hat{s}}$ in second order perturbation theory is
\begin{equation}
\Delta E_i = -\xi^2 \sum_{j \neq i} \frac{\left| \bra{n}  \mathbf{\hat{l}} \cdot \mathbf{\hat{s}} \ket{k} \right|^2}{E_j - E_i},
\label{SOCpert}
\end{equation}
If the unperturbed Hamiltonian commutes with the spin operator, $\ket{i}$ has a well defined spin  $\sigma_i$ but can be considered a superposition of different orbitals $\mu$ so in the simplest case (ignoring other quantum numbers, e.g., $\mathbf{k}$)
\begin{equation}
\ket{i} = \sum_{\mu} c_{i,\mu} \ket{ \mu, \sigma_i}.
\label{multisite_orbs}
\end{equation}  
For d-electron magnetism, which is of focus here, it is suitable to consider $\mu$ as $\text{d}_{z^2}$, $\text{d}_{xz}$, $\text{d}_{yz}$, $\text{d}_{xy}$ or $\text{d}_{x^2-y^2}$. The numerator in Eq.~\ref{SOCpert} then contains matrix elements $\bra{\text{d}_i, \sigma_i} \mathbf{l}\cdot\mathbf{s} \ket{\text{d}_j, \sigma_j}$, which determine the effect of the SOC. For convenience these matrix elements are explicitly listed in Tabel~\ref{SOmatrix}, with $\theta$ and $\phi$ denoting the polar and azimuthal angles of the spin quantization axis relative to the crystal lattice. 
\setlength\tabcolsep{3pt}
\begin{table*} {
\caption{\label{SOmatrix} Matrix elements $\bra{\sigma_i , \text{d}_i } \mathbf{\hat{l}} \cdot \mathbf{\hat{s}}\ket{\sigma_j , \text{d}_j}$ of the spin-orbit coupling operator with respect to spin states in direction $\hat{\mathbf{n}} = (\sin \theta \cos \phi , \sin \theta \sin \phi, \cos \theta)$ and d-orbitals, in units of $\hbar^2$. Reproduced from Ref.~\cite{Abate1965}.}
	\begin{tabular}{  l  c  c  c  c  c }
	\hline	\hline
		     			& $\ket{\uparrow, \text{d}_{xy}}$	& $\ket{\uparrow, \text{d}_{yz}}$  	& $\ket{\uparrow, \text{d}_{z^2}}$ 	& $\ket{\uparrow, \text{d}_{xz}}$ 	& $\ket{\uparrow, \text{d}_{x^2 - y^2}}$ \\ \hline
	$\bra{\uparrow, \text{d}_{xy}}$	& $\scriptstyle 0$	& $\scriptstyle \frac{1}{2}\img\sin{\theta}\sin{\phi}$ & $\scriptstyle 0$ & $\scriptstyle -\frac{1}{2}\img\sin{\theta}\cos{\phi}$ & $\scriptstyle \img\cos{\theta}$	 \\ 
	$\bra{\uparrow, \text{d}_{yz}}$	& -${\scriptstyle \frac{1}{2}\img\sin{\theta}\sin{\phi}}$ & $\scriptstyle 0$			& $\scriptstyle -\frac{\sqrt{3}}{2}\img\sin{\theta}\cos{\phi}$ & $\scriptstyle \frac{\img}{2}\cos{\theta}$ & $\scriptstyle \frac{-\img}{2}\sin{\theta}\cos{\phi}$ \\
	$\bra{\uparrow, \text{d}_{z^2}}$	& $\scriptstyle 0$	& $\scriptstyle \frac{\sqrt{3}}{2}\img\sin{\theta}\cos{\phi}$ & $\scriptstyle 0$		& $\scriptstyle -\frac{\sqrt{3}}{2}\img\sin{\theta}\sin{\phi}$ &  $\scriptstyle 0$		 \\ 
	$\bra{\uparrow, \text{d}_{xz}}$	& $\scriptstyle \frac{1}{2}\img\sin{\theta}\cos{\phi}$ & $\scriptstyle-\frac{\img}{2}\cos{\theta}$ & $\scriptstyle \frac{\sqrt{3}}{2}\img\sin{\theta}\sin{\phi}$ & $\scriptstyle 0$		& $\scriptstyle -\frac{1}{2}\img\sin{\theta}\sin{\phi}$ \\ 
	$\bra{\uparrow, \text{d}_{x^2-y^2}}$	& $\scriptstyle -\img\cos{\theta}$ & $\scriptstyle \frac{-\img}{2}\sin{\theta}\cos{\phi}$ & $\scriptstyle 0$ & $\scriptstyle \frac{1}{2}\img\sin{\theta}\sin{\phi}$ &  $\scriptstyle 0$			 \\ \hline
	$\bra{\downarrow, \text{d}_{xy}}$	& $\scriptstyle 0$	& $\substack{-\frac{1}{2}(\cos{\phi} \\- \img\cos{\theta}\sin{\phi})}$ & $\scriptstyle 0 $ & $\substack{-\frac{1}{2}(\sin{\phi} \\ + \img\cos{\theta}\cos{\phi})}$ & $\scriptstyle -\img\sin{\theta}$	 \\ 
	$\bra{\downarrow, \text{d}_{yz}}$	& $\substack{\frac{1}{2}(\cos{\phi} \\- \img\cos{\theta}\sin{\phi})}$ & $\scriptstyle 0$	& $\substack{-\frac{\sqrt{3}}{2}(\sin{\phi} \\+ \img\cos{\theta}\cos{\phi})}$ & $\scriptstyle -\frac{\img}{2}\sin{\theta}$ & $\substack{-\frac{1}{2}(\sin{\phi} \\+ \img \cos{\theta}\cos{\phi})}$  \\ 
	$\bra{\downarrow, \text{d}_{z^2}}$	& $\scriptstyle 0$	& $\substack{\frac{\sqrt{3}}{2}(\sin{\phi} \\ + \img\cos{\theta}\cos{\phi})}$ & $\scriptstyle 0$ & $\substack{\frac{\sqrt{3}}{2}(\cos{\phi} \\ - \img \cos{\theta}\sin{\phi})}$ & $\scriptstyle 0$ \\ 
	$\bra{\downarrow, \text{d}_{xz}}$	& $\substack{\frac{1}{2}(\sin{\phi} \\ + \img\cos{\theta}\cos{\phi})}$ & $\scriptstyle \frac{\img}{2}\sin{\theta}$ & $\substack{-\frac{\sqrt{3}}{2}(\cos{\phi} \\ - \img\cos{\theta}\sin{\phi})}$ & $\scriptstyle 0$ &  $\substack{\frac{1}{2}(\cos{\phi} \\ - \img\cos{\theta}\sin{\phi})}$	 \\ 
	$\bra{\downarrow, \text{d}_{x^2-y^2}}$	& $\scriptstyle \img\sin{\theta}$ & $\substack{-\frac{1}{2}(\sin{\phi} \\ + \img \cos{\theta}\cos{\phi})}$ & $\scriptstyle 0$ & $\substack{-\frac{1}{2}(\cos{\phi} \\ - \img \cos{\theta}\sin{\phi})}$ &  $\scriptstyle 0$			 \\ \hline\hline
	\end{tabular}	}
\end{table*} \setlength\tabcolsep{6pt}

As mentioned in the main text, only coupling between states $\ket{i}$ and $\ket{j}$ with energies $E_i$ and $E_j$ such that $E_i < E_\text{F} < E_j$ will contribute to the MAE and clearly then $\Delta E_i \leq 0$ according to Eq.~\ref{SOCpert}. In terms of the matrix elements in Table~\ref{SOmatrix} this means that any coupling containing $\cos \theta$ will lower the energy for $\theta=0$, i.e. favoring a uniaxial magnetization (positive MAE), while $\sin \theta$ lowers the energy for $\theta=\pi/2$ which favors in-plane magnetization (negative MAE). The situation taking into account multiple atomic types and hybridisation in Eq.~\ref{HybridisationSOC} is somewhat more complicated and contains a product of matrix elements for possibly different atomic types. Nevertheless, the MAE is still determined by the matrix elements in Table~\ref{SOmatrix}.

\bibliography{literature}{}
\bibliographystyle{apsrev4-1}

\end{document}